\documentclass[reprint,twocolumn,showpacs,superscriptaddress,longbibliography,email,floatfix,aps,pra]{revtex4-2}

\usepackage{header}
\usepackage[normalem]{ulem}
\newcommand{\figref}[1]{Fig.\,\ref{#1}}
\newcommand{\subfigref}[2]{Fig.\,\hyperref[#1]{#2}}

\newcommand{\appref}[2]{Appendix \hyperref[#1]{#2}}
\renewcommand{\eqref}[1]{Eq.\,(\ref{#1})}

\newcommand{\multieqref}[2]{Eqs.\,(\ref{#1} - \ref{#2})}
\usepackage{amssymb,amsmath,mathtools}
\usepackage{physics} 
\usepackage{relsize} 
\usepackage{dsfont}
\usepackage{bm} 
\usepackage{amsthm}
\usepackage{svg}
\newcommand{\mycaption}[2]{\caption[#1]{\textbf{#1} #2}}
\DeclareMathOperator{\Var}{Var}

\usepackage{braket}
\newcommand{\superket}[1]{|#1\rangle\rangle}
\newcommand{\superbra}[1]{\langle \langle#1|}

\begin{document}
\title{The Boundary Time Crystal as a light source for collectively enhanced sensing}
\author{Malik Jirasek}
\affiliation{Institut f\"ur Theoretische Physik, Universit\"at T\"ubingen, Auf der Morgenstelle 14, 72076 T\"ubingen, Germany}
\author{Igor Lesanovsky}
\affiliation{Institut f\"ur Theoretische Physik, Universit\"at T\"ubingen, Auf der Morgenstelle 14, 72076 T\"ubingen, Germany}
\affiliation{School of Physics and Astronomy and Centre for the Mathematics and Theoretical Physics of Quantum Non-Equilibrium Systems, The University of Nottingham, Nottingham, NG7 2RD, United Kingdom}
\author{Albert Cabot}
\affiliation{Institut f\"ur Theoretische Physik, Universit\"at T\"ubingen, Auf der Morgenstelle 14, 72076 T\"ubingen, Germany}
\affiliation{Institute for Cross-Disciplinary Physics and Complex Systems (IFISC) UIB-CSIC, Campus Universitat Illes Balears, 07122, Palma de Mallorca, Spain.}

\begin{abstract}
    Modern precision measurements, such as interferometry for detecting gravitational waves, rely on the estimation of optical phases encoded in light fields. Here, we propose to exploit the collectively enhanced output field of a driven-dissipative many-body quantum system as a light source in order to improve the precision of estimating optical phases. Pronounced temporal correlations of such output fields benefit the sensitivity of measurement protocols, which we show theoretically by employing a boundary time crystal  as a light source. 
    The fundamental bound on the precision of such estimation shows scaling with the number of constituents $N$ of the many-body system as $N^4$ while scaling linearly with the measurement time $T$. We discuss this scaling both from a perspective of the resources employed to build the light source and of the resources produced by the light source. We show that a measurement scheme, in which the phase shifted light field is guided into an auxiliary replica system, which serves as a detector that is sensitive to non-trivial temporal correlations of light, can  saturate the fundamental bound on precision at an optimal operating point.
\end{abstract}

\maketitle


\section{Introduction}

Quantum technologies promise sensors that surpass the precision of their traditional counterparts, for example in magnetometry \cite{Budker2007} or interferometry for the detection of gravitational waves \cite{Aasi2013}. Sensing applications might also benefit from novel light sources that rely on the exploitation of collective phenomena in driven-dissipative quantum systems \cite{Meiser2009superradiant_laser,meiser2010intensity,martin2011extreme,westergaard2015observation}. These may feature steady state superradiance \cite{Meiser2009superradiant_laser,meiser2010intensity}, synchronization \cite{Xu2014collective,Xu2015collective_ramsey,cabot2021synchronization}, or time crystal phases \cite{Iemini2018,Cabot2023,Kessler2021}, that influence the properties of their light emission. In the case of time crystal phases, their characteristic asymptotic oscillatory behavior \cite{Sacha2018, Else2020, Zaletel2023} leads to persistent temporal correlations in their emitted light \cite{Carmichael1980, Cabot2023}. These oscillations occur either with a period that is a discrete multiple of that of an external driving \cite{Else2016, Khemani2016, Gong2018, Yao2017, Wang2018, Gambetta2019, Lazarides2020, Riera-Campeny2020, Zhu2019, Chinzei2020, Kessler2021, Tuquero2022, Sarkar2022, Cabot2022,Yousefjani2025nonhermitian}, or with a period varying continuously with system parameters, in the case of time-independent driving \cite{Iemini2018, Carollo2022, Tucker2018, Lledó2020,  Buca2019, Buca2019_1, Booker2020, Buonaiuto2021, Prazeres2021, Piccitto2021, Lourenco2022, Hajdusek2022, Krishna2023, Mattes2023}. 
An example of the latter case is the boundary time crystal (BTC), emerging in driven-dissipative collective spin systems \cite{Iemini2018, Carollo2022, Cabot2023}. In its stationary regime, the light output displays the same statistics as the laser driving it, while in its time crystal regime, it exhibits intricate time-correlations which become more pronounced with system size \cite{Carmichael1980, Cabot2023,somech2024quantum}.

\begin{figure}[t]
    \label{fig:measurement_setups}
    \centering
    \includegraphics{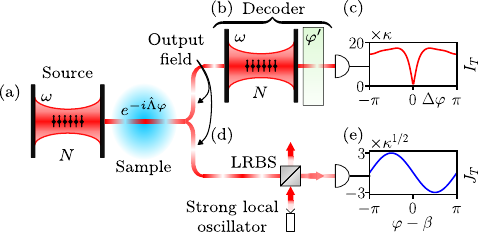}
    \vspace{-0.4cm}
    \mycaption{The boundary time crystal as a light source.}{(a) Light emitted by a boundary time crystal (BTC) composed of $N$ two-level systems driven with Rabi frequency $\omega$ is guided through a sample. A phase shift $\varphi$ is encoded in the light, described by the unitary $e^{-i\hat{\Lambda}\varphi}$. $\hat{\Lambda}$ is the photon count operator. (b) Phase shifted source light is cascaded into a decoder BTC with additional tunable phase shift $\varphi'$. (c) Time-averaged intensity output by the source-decoder system $I_T$ as a function of $\Delta \varphi = \varphi - \varphi'$ for $N=6$ and $\omega/\omega_\mathrm{c}=4$. (d) Phase shifted source light is superposed with the field of a strong local oscillator in a homodyne detection scheme using a low reflectivity beam splitter (LRBS) with homodyne phase $\beta$. (e) Time-averaged homodyne current measured at the indicated output port as a function of $\varphi-\beta$ for $N=6$ and $\omega/\omega_\mathrm{c} = 0.5$.}
    \vspace{-0.6cm}
\end{figure}
 
In this paper we demonstrate how driven-dissipative many-body quantum systems can be exploited as a source of light that enables  collectively enhanced parameter estimation. We showcase the advantage of this method by investigating the setting depicted in \subfigref{fig:measurement_setups}{1(a)}, in which the output light of a BTC is used to measure an external phase shift $\varphi$ imprinted by a sample. 
Our results reveal that time crystal phases can be exploited more efficiently by using their emitted light for measuring external parameters, e.g. phase shifts, compared to recent studies on the sensing of internal parameters of time crystals \cite{Montenegro2023,Cabot2024, iemini2024floquet, Cabot2025, Mattes2025,Albarelli2025,gribben2025boundary,Yousefjani2025discrete,Yousefjani2025discretetime}. This advantage is quantified by the quantum Fisher information (QFI) of the source system and its output field, which defines the fundamental precision with which the emitted light permits determination of the phase shift \cite{Gammelmark2014,Macieszczak2014}. We analyze the resources governing the sensitivity bounds for the estimation of $\varphi$ from the light emitted by a BTC, focusing on the scaling of the QFI with the number of emitters $N$ and measurement time $T$. We find that the time crystal phase provides the most advantageous scaling, with the QFI growing as $N^4 T$, while in the stationary regime it grows as $N^2 T$.

In order to tap the  enhancement anticipated by the QFI, we propose two different measurement protocols to estimate
$\varphi$ from the BTC output light, and analyze their corresponding Fisher information (FI). The first protocol is akin to homodyne detection \cite{Ilias2022, Gambetta2001, Catana2012,gammelmark_bayesian_2013,Kiilerich2014,Xu2015collective_ramsey, Albarelli2017, Albarelli2018, Shankar2019, Albarelli2020, Rossi2020,amoros2021noisy, Nurdin2022,radaelli_parameter_2024,amoros2025noisy, Mattes2025} [see \subfigref{fig:measurement_setups}{1(d)-(e)}], while in the second protocol, the emitted light of a BTC is guided into an auxiliary replica system acting as a perfect absorber for the phase shifted light \cite{Yang, Godley2023} [see \subfigref{fig:measurement_setups}{1(b)-(c)}]. 
We show that the FI of both protocols saturate the QFI in the stationary regime, 
while this is only achieved by the second measurement protocol in the time crystal regime.

\section{Fundamental bounds on precision}

To set the stage, we briefly introduce the fundamental bound on precision with which an external phase shift can be measured, using a dissipative many-body quantum system as a light source. The dynamics shall be described by a Lindblad master equation ($\hbar = 1$)
\begin{align}
    \partial_t \rho = -i[\hat{H},\rho] + \kappa\mathcal{D}[\hat{L}]\rho =:\mathcal{L}[\rho],
    \label{eq:lindblad master equation}
\end{align}
with the dissipator $\mathcal{D}[\mathcal{O}]\rho=\mathcal{O}\rho \mathcal{O}^\dagger - \{\mathcal{O}^\dagger \mathcal{O} , \rho \}/2$ and the reduced system state $\rho$. The Hamiltonian $\hat{H}$ describes coherent processes, e.g.,  coherent driving of the source system by the light inside a cavity [cf. \subfigref{fig:measurement_setups}{1(a)}]. The jump operator $\hat{L}$ describes photon emission by the source system at rate $\kappa$, e.g., out of the cavity into a waveguide. In the input-output formalism \cite{ZollerGardiner}, the global state of system and environment is a matrix product state (MPS) $\ket{\psi_\varphi}$ \cite{Gammelmark2014,Macieszczak2014,Yang}. When guided through an unknown sample, the light emitted by the source system acquires a phase shift $\varphi$ [see \subfigref{fig:measurement_setups}{1(a)}], which is described by the unitary transformation $\ket{\psi_\varphi} = e^{-i\hat{\Lambda}\varphi} \ket{\psi_0}$ \cite{Macieszczak2014}. Here the state $\ket{\psi_0}$ is the global state in absence of a phase shift, and $\hat{\Lambda}$ is the photon count operator (see Appendix \ref{SecSM:QFI} for details). Effectively, the unitary transformation of the global state can be expressed as a transformation of the system jump operator corresponding to the emission of a photon $\hat{L}\to e^{-i\varphi}\hat{L}$ \cite{Macieszczak2014}. 

As a figure of merit for the performance of a measurement of the parameter $\varphi$, we consider the estimation error $\delta \varphi$. This is formally given by the standard deviation of an unbiased estimator specific for a given measurement protocol \cite{Cramér1946, Braunstein1994, WisemanMilburn2009, Cabot2024}. Optimizing over all possible measurements yields the quantum Cram\'er-Rao bound (QCRB) for the estimation error \cite{Braunstein1994, WisemanMilburn2009}
\begin{align}
    \delta\varphi \geq \frac{1}{\sqrt{F_\varphi(\varphi_0,T)}},
    \label{eq:qcrb main}
\end{align}
where $F_\varphi(\varphi_0,T)$ is the global QFI of the system and environment state for estimating $\varphi$, evaluated at the actual parameter value $\varphi_0$ and for a total measurement time $T$ \cite{Gammelmark2014,Macieszczak2014}. Although this differs in principle from the QFI of the environment alone, for long measurement times the contribution by the environment scales linearly with $T$ while the system contribution does not  \cite{Yang}. Thus, in this limit the relative difference between the global and the environment QFI vanishes. Below, we also consider the FI of the homodyne or photocounting records, $\mathcal{I}_\varphi^\mathrm{(u)}(\varphi_0,T)$ with $\mathrm{u=h,c}$, respectively, which bound estimation protocols based solely on these monitoring records \cite{gammelmark_bayesian_2013,Kiilerich2016,Albarelli2018,Kiilerich2016}. These FI are generally smaller than $F_\varphi(\varphi_0,T)$  \cite{Gammelmark2014,Kiilerich2016,Albarelli2018,radaelli_parameter_2024}, though continuous-monitoring protocols can be optimal and saturate the QFI \cite{Mattes2025}. 

For long measurement times, one finds linear growth of the QFI with $T$, such that the QFI rate $f_\varphi := \lim_{T\to\infty} F_\varphi(\varphi_0,T) /T$ only depends on the system parameters and the estimated phase shift $\varphi$. This linear scaling arises from the finite coherence time of the system at finite system sizes $N$, see also \cite{Macieszczak2014}. By employing the MPS description of the global system and environment state we further find the long-time QFI rate (see Appendix \ref{SecSM:QFI})
\begin{align}
    \frac{f_\varphi}{4\kappa} = &\frac{2\kappa}{T}\int_0^T \!\! dt \int_0^{T-t} \hspace{-0.5cm} d\tau \;C(\tau) +\langle \hat{L}^\dagger \hat{L}\rangle_\mathrm{ss} - \kappa T \langle \hat{L}^\dagger \hat{L}\rangle_\mathrm{ss}^2,
    \label{eq:long-time qfi rate}
\end{align}
where $C(\tau) := \Tr [ \hat{L}^\dagger \hat{L} e^{\mathcal{L}\tau}(\hat{L} \rho_\mathrm{ss} \hat{L}^\dagger) ]$ and $\langle \mathcal{O}\rangle_\mathrm{ss} = \Tr [\mathcal{O}\rho_\mathrm{ss}]$ with the stationary state $\rho_\mathrm{ss}$. This shows that the QFI rate is given by the time integrated two-time intensity correlation of the system output \cite{WisemanMilburn2009, Macieszczak2014} and independent of the actual parameter value $\varphi_0$. 

\section{The Boundary Time Crystal as a light source}

In the following, we consider the BTC as the light source.
The BTC is composed of $N$ two-level emitters with resonant driving with Rabi frequency $\omega$ and collective dissipation with rate $\kappa$, as described by the Hamiltonian $\hat{H}=\omega\hat{S}_\mathrm{x}$ and the jump operator $\hat{L} = \hat{S}_-$, respectively. The operators $\hat{S}_\alpha = \sum_{i=1}^N \hat{\sigma}_\alpha^i /2$, with $\alpha\in\{\mathrm{x,y,z}\}$, are collective total angular momentum operators, where the Pauli matrices $\hat{\sigma}_\alpha^i$ correspond to emitter $i$, and the ladder operators are $\hat{S}_\pm = \hat{S}_\mathrm{x}\pm i\hat{S}_\mathrm{y}$.
The BTC features a phase transition in the thermodynamic limit $(N\to\infty)$ from a stationary regime to a time crystal regime. The transition occurs at the critical Rabi frequency $\omega_\mathrm{c} = N\kappa/2$ \cite{Carmichael1980}, which implies that Rabi frequencies have to be adjusted with system size when considering fixed values of $\omega/\omega_\mathrm{c}$.  For $\omega < \omega_\mathrm{c}$, the system displays a stationary regime and rapidly approaches a stationary state via overdamped relaxation \cite{Carmichael1980, Cabot2023}. For $\omega > \omega_\mathrm{c}$, the system is in a time crystal regime, characterized by oscillating system observables \cite{Carmichael1980, Cabot2023}. Beyond time crystals, this and related models have also been explored in the context of cooperative fluorescence and driven superradiance \cite{Carmichael1980,martin2011extreme,Hannukainen2018,Ferioli2023,somech2024quantum,agarwal_directional_superradiance,goncalves_phase_separation}.

We determine the bound on the phase estimation error $\delta\varphi$ with a BTC as light source by computing the QFI rate \eqref{eq:long-time qfi rate} in both regimes. In the stationary regime, a Holstein-Primakoff (HP) approach yields $f_\varphi \approx 4\omega^2/\kappa$ (see Appendix \ref{SecSM:HP approach single system}), which for constant $\omega/\omega_\mathrm{c}$ yields a  $N^2$ scaling of the QFI rate.
Deep in the time crystal regime ($\omega/\omega_\mathrm{c} \gg 1$), we obtain $C(\tau)$ analytically by using the superspin method \cite{Nemeth2025, Albarelli2025}, resulting in the long-time QFI rate (see Appendix \ref{SecSM:superspin})
\begin{align}
     f_\varphi \approx \kappa N(N+2)\left[\frac{(N-1)(N+3)}{135} + \frac{2}{3}\right] =:f_{\varphi,\infty} ,
    \label{eq:infinite driving qfi}
\end{align}
where the approximation becomes exact when $\omega/\omega_\mathrm{c}\to\infty$. This shows that the sensitivity of the BTC light for measuring optical phases is enhanced by pronounced temporal intensity correlations of the output field in the time crystal regime. These correlations grow with the number of emitters in the BTC \cite{Carmichael1980}, leading to a scaling of $f_{\varphi,\infty} \propto N^4$ for large, but finite system sizes. See \ref{SecSM:superspin} and Appendix \ref{SecSM:HP approach single system} for a numerical benchmark of these analytical expressions.

It is insightful to examine the output state of light of the BTC in both of its regimes and its influence on the QFI. In the stationary regime, the output field is in a coherent state \cite{somech2024quantum,Cabot2024}, and therefore offers no metrological advantage over the input driving field. This reflects the absence of photon correlations in the output, which causes the first and last terms in \eqref{eq:long-time qfi rate} to cancel.  The increase of the QFI with system size $N$ stems from the fact that the driving strength $\omega$ (set by the input‑light power) must scale with $N$ to maintain the same operating point in the phase diagram. Thus in the stationary regime, the improvement in precision arises solely from using more light for larger $N$. In the time crystal regime, however, the emitted light exhibits correlations,  reflected in oscillating two‑time intensity correlations whose strength and quality factor increases with system size (see Appendix \ref{SecSM:superspin}). These increasingly strong temporal correlations provide an additional metrological resource, leading to the extra $N^2$ scaling of the QFI in this phase. This resource is distinct from pure increase of driving intensity with larger systems and provides a collective enhancement in sensing optical phases.


\begin{figure}
    \centering
    \vspace{-0.3cm}
    \includegraphics{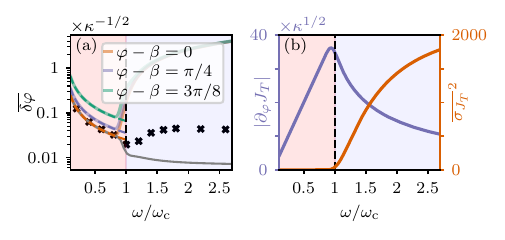}
    \vspace{-0.6cm}
    \mycaption{Results for the average homodyne current protocol.}{(a) Time rescaled long-time estimation error $\overline{\delta\varphi}$ for the homodyne protocol versus Rabi frequency $\omega$, for $N=40$ and various $\varphi-\beta$. Dashed lines: corresponding HP approximated curves. Black crosses: bound given by the FI $[\mathcal{I}^{(h)}_\varphi]^{-1/2}$ averaged over 1000 trajectories (see Appendix \ref{SecSM:classical_FI} for details). Gray line: bound given by the QFI $f_\varphi^{-1/2}$. (b) Derivative and variance of the long-time averaged homodyne current versus Rabi frequency for $N=40$ and $\varphi-\beta = 0$.}
    \label{fig:estimation error homodyne protocol}
    \vspace{-0.3cm}
\end{figure}

\section{Homodyne current protocols}

Whether the minimum estimation error given by \eqref{eq:qcrb main} can be practically achieved depends on the actual measurement performed on the BTC light and the processing of measurement outputs. In homodyne protocols, the source output light first probes the unknown sample acquiring the phase shift $\varphi$ and it is then superposed with the emission field of a strong local oscillator [see \subfigref{fig:measurement_setups}{1(d)}]. This setup allows for the measurement of the homodyne current extracted at the indicated output port \cite{WisemanMilburn2009, Cabot2023}
\begin{align}
    J_\beta (t) = \sqrt{\kappa}\langle \hat{S}_\beta (t)\rangle_\mathrm{H} + \frac{dW}{dt},
    \label{eq:homodyne current}
\end{align}
where $\hat{S}_\beta = \hat{S}_- e^{i(\beta-\varphi)}+\hat{S}_+e^{-i(\beta-\varphi)}$ is a general quadrature, with the homodyne phase $\beta$ and $dW/dt$ the derivative of a Wiener process. The subscript $\mathrm{H}$ indicates expectation values with respect to the homodyne conditional state \cite{WisemanMilburn2009, Cabot2023}.

We assume ideal monitoring of the homodyne current over a long measurement time $T$, and neglect any additional loss channels.
We first consider the protocol where $\varphi$ is estimated from the time average of the homodyne current, $J_T = (1/T)\int_0^T J_\beta (t) dt$, and analyze its asymptotic estimation error for long measurement times. Afterwards, we also study the FI of the full homodyne record, $\mathcal{I}_{\varphi}^\mathrm{(h)}(\varphi_0,T)$, to determine how much of the QFI can be in principle extracted by this kind of continuous measurement. The average current reads $\mathbb{E}[J_T] = (\sqrt{\kappa}/T)\int_0^T \langle \hat{S}_\beta (t)\rangle dt$ with $\mathbb{E}[\bullet]$ the average over trajectories and $\langle \mathcal{O}(t) \rangle = \Tr [\mathcal{O}\rho(t)]$ the unconditional expectation value. For the estimation error, the error propagation formula holds \cite{Cabot2024}
\begin{align}
    \delta \varphi = \sqrt{\mathbb{E}[J_T^2] - \mathbb{E}[J_T]^2}\left| \frac{\partial \mathbb{E}[J_T]}{\partial\varphi} \right|^{-1},
    \label{eq:error propagation homodyne}
\end{align}
which is lower bounded by the QCRB \eqref{eq:qcrb main}. For long measurement times, the time averaged current approaches its ensemble mean $\lim_{T\to\infty} J_T = \sqrt{\kappa}\langle\hat{S}_\beta \rangle_\mathrm{ss}$ \cite{WisemanMilburn2009}. Far away from critical points, large deviations theory yields a behavior of the standard deviation as $\sqrt{\mathbb{E}[J_T^2] - \mathbb{E}[J_T]^2} \sim  \overline{\sigma_{J_T}}/\sqrt{T}$ \cite{Hickey}, with the time independent rescaled standard deviation $\overline{\sigma_{J_T}}$. This implies a behavior of the estimation error as $\delta \varphi \sim \overline{\delta\varphi}/\sqrt{T}$, with the long-time rescaled estimation error $\overline{\delta\varphi}$. The HP approach in the stationary regime yields (see Appendix \ref{SecSM:largedeviations} and \ref{SecSM:HP approach single system})
\begin{align}
    \overline{\delta\varphi} \approx \frac{\sqrt{\kappa}}{2\omega|\cos(\varphi-\beta)|}.
    \label{eq:estimation error homodyne stationary}
\end{align}
This agrees with the numerical results presented in \subfigref{fig:estimation error homodyne protocol}{2(a)} deep in the stationary regime, and saturates the QCRB for $\varphi-\beta = 0$. For $\varphi-\beta = 0$, the average homodyne current vanishes, which can be interpreted as a signature in the measurement data when tuning through the parameter $\beta$ [see \subfigref{fig:measurement_setups}{1(e)}]. In the vicinity of the critical Rabi frequency, the long-measurement-time estimation error increases to a higher level in the time crystal regime than the minimum observed in the stationary regime, due to a larger variance and smaller susceptibility of the homodyne current [see \subfigref{fig:estimation error homodyne protocol}{2(b)}].

In \subfigref{fig:estimation error homodyne protocol}{2(a)}, we also show  the error bound (rate) for estimation protocols based on the homodyne current, given by $[\mathcal{I}^{\mathrm{(h)}}_\varphi]^{-1/2}$ (crosses), with $\mathcal{I}^{\mathrm{(h)}}_\varphi := \lim_{T\to\infty} \mathcal{I}^{\mathrm{(h)}}_\varphi(\varphi_0,T) /T$. The minimum achievable error is near the transition, while for the time crystal phase the error bound displays similar values as the largest ones observed in the stationary phase. Protocols that exploit temporal correlations in the homodyne current \cite{martin2011extreme,Kiilerich2016} may outperform the average‑current method in this regime. However, for the system sizes and parameter regimes studied (see Appendix \ref{SecSM:classical_FI}), the $N^4$ scaling of the QFI does not emerge in $\mathcal{I}^{\mathrm{(h)}}_\varphi$. The reason is that the light emitted by the local oscillator is in a coherent state and its superposition with the BTC light is therefore not sufficient to decode the intricate correlations leading to the enhancement of the QFI.


\section{Perfect absorber protocol}
 
We now consider the measurement protocol developed in Refs. \cite{Yang, Godley2023}, which requires constructing a perfect absorber for the BTC light \cite{Stannigel2012}. In fact, this absorber is a replica BTC acting as a 'decoder' for the information inscribed on the phase shifted light [see \subfigref{fig:measurement_setups}{1(b)}].

\begin{figure}
    \centering
    \vspace{-0.5cm}
    \includegraphics{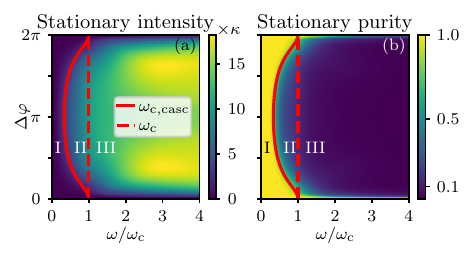}
    \vspace{-0.6cm}
    \mycaption{Stationary properties of the cascaded system.}{$N=6$ for all panels. (a) Emitted intensity and (b) purity of the stationary state varying Rabi frequency $\omega$ and phase difference $\Delta \varphi$. Red-solid lines: mean-field transition line $\omega_\mathrm{c, casc}$ from the stationary regime (\MakeUppercase{\romannumeral 1}) to the intermediate regime, where only the decoder is in a time crystal regime (\MakeUppercase{\romannumeral 2}). Red-dashed lines: transition line $\omega_\mathrm{c}$  where both source and decoder display time crystal behavior (\MakeUppercase{\romannumeral 3}).}
    \label{fig:int pur ent stationary casc}
    \vspace{-0.3cm}
\end{figure}

The light emitted by the source BTC acquires a phase $\varphi$ and is then unidirectionally guided into the decoder system. By applying an auxiliary tunable phase shift $\varphi'$ to the decoder output, the phase $\varphi$ can be estimated by means of an intensity measurement of the decoder output [see \subfigref{fig:measurement_setups}{1(c)}]. The evolution of the emerging cascaded system shown in \subfigref{fig:measurement_setups}{1(b)} is described by the master equation \eqref{eq:lindblad master equation} with Hamiltonian $\hat{H} = \omega(\hat{S}_\mathrm{x}^\mathrm{S} + \hat{S}_\mathrm{x}^\mathrm{D})- (i\kappa /2) (e^{-i\Delta\varphi} \hat{S}_+^\mathrm{D}\hat{S}_-^\mathrm{S} - e^{i\Delta\varphi} \hat{S}_+^\mathrm{S}\hat{S}_-^\mathrm{D})$, jump operator $\hat{L} = \hat{L}_\mathrm{casc} = e^{-i\Delta\varphi} \hat{S}_-^\mathrm{S} + \hat{S}_-^\mathrm{D}$, the phase difference $\Delta\varphi = \varphi - \varphi'$, and the superscript $\mathrm{S}$ ($\mathrm{D}$) labels the collective operators of the source (decoder) \cite{Stannigel2012, Yang, Cabot2024}. The cascaded system features a pure dark state at $\Delta \varphi =0$ \cite{Cabot2024}, which renders the decoder system a perfect absorber for this choice of parameters \cite{Stannigel2012}. The occurrence of the dark state is independent of $\omega$, as indicated in \subfigref{fig:int pur ent stationary casc}{3(a)}. For general values of $\Delta\varphi$, the behavior with respect to the Rabi frequency displays three different regimes. For $\omega < \omega_\mathrm{c,casc}=\sqrt{5-4\cos (\Delta\varphi)}$, both source and decoder are in the single system stationary regime (regime \MakeUppercase{\romannumeral 1} in \figref{fig:int pur ent stationary casc}). For $\omega_\mathrm{c, casc} \leq \omega < \omega_\mathrm{c}$, the decoder first transitions to a time crystal regime, with its observables showing persistent oscillations in the thermodynamic limit (\MakeUppercase{\romannumeral 2}) (see Appendix \ref{SecSM:HP absorber}). For $\omega \geq \omega_\mathrm{c}$, both source and decoder are in a time crystal regime (\MakeUppercase{\romannumeral 3}). These regimes are also characterized by their subsystem entropies, which we show in Appendix \ref{SecSM:HP absorber}.

\begin{figure*}[t!]
    \centering
    \vspace{-0.3cm}
    \includegraphics[width=\linewidth]{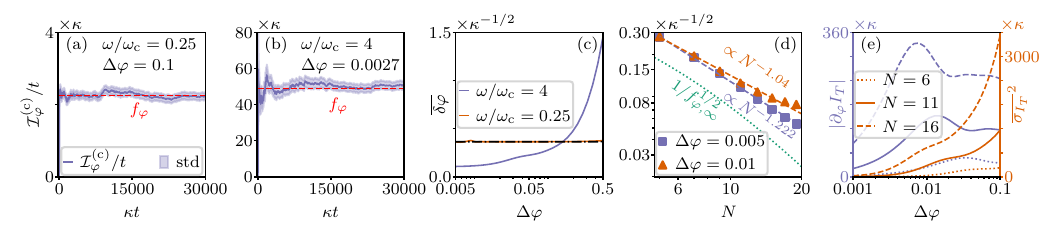}
    \vspace{-0.8cm}
    \mycaption{Results for the perfect absorber protocol.}{(a–b) Fisher information (FI) rate for photocounting in the perfect absorber protocol with $N=6$: (a) stationary regime ($\omega/\omega_{\mathrm{c}}=0.25$, $\Delta\varphi=0.1$); (b) time crystal regime ($\omega/\omega_{\mathrm{c}}=4$, $\Delta\varphi=0.0027$). Purple solid line: FI averaged over 1000 trajectories, with the shaded region indicating the Monte-Carlo standard deviation. Red-dashed line: corresponding long-time QFI rate. (c) Time-rescaled long-time estimation error versus $\Delta\varphi$ for $N=11$, comparing the cascaded time crystal regime ($\omega/\omega_{\mathrm{c}}=4$, \MakeUppercase{\romannumeral 3}) and the cascaded stationary regime ($\omega/\omega_{\mathrm{c}}=0.25$, \MakeUppercase{\romannumeral 1}). Black dash-dotted line: HP approximation of the error for $\omega/\omega_{\mathrm{c}}=0.25$. (d) Estimation error versus system size for different $\Delta\varphi$ at $\omega/\omega_{\mathrm{c}}=4$. Green-dotted line:  QFI rate in the limit $\omega/\omega_{\mathrm{c}}\to\infty$ [cf. \eqref{eq:infinite driving qfi}]. Dashed lines: power-law fit $\overline{\delta\varphi}=bN^{-\alpha}$ on the data. (e) Derivative and variance of the time-averaged intensity versus $\Delta\varphi$ for various system sizes with $\omega/\omega_{\mathrm{c}}=4$.}
    \label{fig:est_err_cascaded}
    \vspace{-0.3cm}
\end{figure*}

In order to assess the potential for measuring the phase shift $\varphi$ using the perfect absorber protocol, we first study the FI rate of the full record of emissions $\mathcal{I}^\mathrm{(c)}_\varphi := \lim_{T\to \infty} \mathcal{I}_\varphi^\mathrm{(c)}(\varphi_0 , T) / T$ [see \subfigref{fig:est_err_cascaded}{4(a),(b)}]. We perform our analysis focusing on a small system size $N=6$, finding that for long measurement times, the FI rate approaches the long-time QFI rate in both regimes. Notice that, due to the low emission intensity near the dark state condition, reaching the stationary FI rate takes a longer time compared to the previous protocols. In the stationary regime, the sensitivity shows little dependence on  $\Delta \varphi$ as long as it is small. In contrast, in the time crystal regime, there is a stronger dependence on  $\Delta \varphi$ and saturation of the QFI occurs when tuning $\Delta \varphi$ to an optimal point (see Appendix \ref{SecSM:classical_FI}).

We further illustrate the performance of this protocol by considering a specific estimation strategy based on the time averaged intensity $I_T = (1/T)\int_0^T dN(t)$ of the cascaded system. $dN(t)$ is a random variable taking the values $1$ if a photon is detected at time $t$ and $0$ otherwise, with average value given by $\mathbb{E}[dN(t)] = \kappa dt\langle\hat{L}_\mathrm{casc}^\dagger \hat{L}_\mathrm{casc} (t)\rangle$ \cite{WisemanMilburn2009}. As in the homodyne case we focus on characterizing the estimation error in the asymptotic limit of long measurement times, in which $\lim_{T\to\infty} I_T = \kappa \langle \hat{L}_\mathrm{casc}^\dagger \hat{L}_\mathrm{casc} \rangle_\mathrm{ss}$, and the estimation error scales as $\delta \varphi \sim \overline{\delta \varphi}/\sqrt{T}$, with $\overline{\delta\varphi} = \overline{\sigma_{I_T}} / |\partial_\varphi I_T|$. Using an HP approach we determine the estimation error in the cascaded stationary regime (see Appendix \ref{SecSM:HP absorber})
\begin{align}
    \overline{\delta \varphi} \approx \sqrt{\frac{\kappa(1-\cos(\Delta\varphi))}{2\omega^2\sin^2(\Delta\varphi)}}. 
    \label{eq:estimation error cascaded stationary}
\end{align}
This saturates the QCRB for $\Delta\varphi \to 0$ and is numerically verified in \subfigref{fig:est_err_cascaded}{4(c)}. At $\Delta \varphi = 0$ the emitted intensity vanishes, which leaves a clear signature in the data when tuning through the parameter $\varphi'$ [see \subfigref{fig:measurement_setups}{1(c)}] and leads to a particularly high sensitivity in the stationary regime. For small but nonzero $\Delta\varphi$, the emitted intensity remains finite and the protocol functions reliably with sensitivity close to its maximum value [see \subfigref{fig:est_err_cascaded}{4(a),(c)}]. In Appendix \ref{SecSM:Construction of the optimal decoder in the stationary regime} we additionally show explicitly how the perfect absorber protocol emerges from the construction in \cite{Yang} in the stationary regime. In the time crystal regime and nearby the perfect absorber condition $\Delta \varphi = 0$, we observe lower values of the estimation error than in the stationary regime, due to a decrease in the variance $\overline{\sigma_{I_T}}^2$ and a maximum in the susceptibility $|\partial_\varphi I_T|$ of the long-measurement-time average intensity. This maximum moves closer to $\Delta\varphi=0$ with increasing system size. Note, however, that the optimal operating point in the time crystal regime is not directly at the dark state condition, but at values of $\Delta \varphi$ slightly above this condition \cite{Yang,Godley2023,Girotti2024}. As a function of system size, we find scaling proportional to $\overline{\delta\varphi} \propto N^{-\alpha}$ with $\alpha=1.222\pm0.018$ for $\Delta\varphi=0.005$, and $\alpha=1.04\pm0.04$ for $\Delta\varphi=0.01$, both for a Rabi frequency of $\omega/\omega_\mathrm{c}=4$ [see \subfigref{fig:est_err_cascaded}{4(d)}]. This analysis shows that while being ideal in the stationary regime, in the time crystal regime estimators based on the long-measurement-time-averaged intensity are not optimal and therefore future work should investigate whether strategies such as maximum likelihood estimation (MLE) \cite{Yang,Godley2023} or pattern counting \cite{Girotti2024} can saturate the FI rate. 


\section{Conclusions and Outlook}

We showed that a BTC can be used as a light source for collectively enhanced sensing of an optical phase shift encoded by an unknown sample. We analyzed fundamental sensitivity bounds through a study of the QFI and evaluated the classical FI of specific measurement strategies based on continuous monitoring. In the time crystal regime, we found that the QFI for sensing optical phases scales as $N^4$ in the size $N$ of the BTC, surpassing the scaling with $N^2$ in the stationary regime. This collective enhancement in sensitivity arises from the correlations imprinted on the emitted light. We have designed a detector based on implementing a perfect absorber and measuring its output via photon counting, showing that the FI of this specific measurement can saturate the QFI of the BTC light for phase estimation.

The largely analytically tractable BTC serves as a convenient theoretical platform for demonstrating the concept behind collectively enhanced parameter estimation using quantum many-body systems as light sources.  
Bridging the gap between this model and practical experimental implementations necessitates the assessment of the effects of finite detection efficiencies, and local noise and decay channels \cite{Albarelli2025,Cabot2025}. While the former have shown to reduce the achievable precision by a factor of $N^2$ in the case of sensing the Rabi frequency of the BTC \cite{Albarelli2025}, it would be an interesting avenue to determine the impact on the sensing scheme presented in our work. 
Beyond the analysis of sensitivity bounds and specific measurement strategies, identifying suitable data processing procedures capable of achieving the reported sensitivities remains an important open question. A natural next step is to explore estimation strategies such as MLE \cite{Yang,Godley2023} or pattern‑counting estimators \cite{Girotti2024}, which have proven effective in other settings implementing the perfect absorber protocol.Another intriguing direction will be to explore whether the observed collective enhancements persist in light emitted from other types of time crystals, particularly in systems involving particles with finite-range interactions and non-collective dissipation \cite{russo2025quantum, wang2025boundary}.


\acknowledgments
\textit{Acknowledgements.---} We acknowledge the use of the QuTiP library \cite{qutip,qutip2,qutip5}. We acknowledge support from the Deutsche Forschungsgemeinschaft (DFG, German Research Foundation) through the Walter Benjamin programme, Grant No. 519847240 and the Research Unit FOR 5413/1. AC acknowledges support from both the Spanish Ministerio de Ciencia, Innovación y Universidades and  Universitat de les Illes Balears through the Beatriz
Galindo programme (BG24/00134). This work was supported by the QuantERA II programme (project CoQuaDis, DFG Grant No. 532763411) that has received funding from the EU H2020 research and innovation programme under GA No. 101017733. This work is also supported by the ERC grant OPEN-2QS (Grant No. 101164443, https://doi.org/10.3030/101164443).

\section*{Data availability and tools}

Code and data in support of the findings of this paper are available on GitHub \cite{github}. Claude Opus v4.7 (Anthropic) was used to assist with editing and refactoring code used in calculations (formatting, variable renaming, suggested algorithmic changes); all suggestions were reviewed, tested, and finalized by the authors.\\


\appendix 

\section{QFI with MPS description}\label{SecSM:QFI}

We consider the estimation of an optical phase shift $\varphi$ imprinted on the photons emitted by a quantum system as described by Eq. (1). This phase shift is described by the transformation of the jump operator $\hat{L} \to \hat{L} e^{-i\varphi}$ \cite{Macieszczak2014}. The estimation error of any measurement of a classical parameter imprinted on the state of a quantum system is lower bounded by the QCRB \cite{Cramér1946, WisemanMilburn2009}
\begin{align}
    \Var[\hat{\varphi}]_{\varphi_0} \geq 1/F_\varphi(\varphi_0, T),
    \label{eq:qcrb}
\end{align}
where $\Var[\hat{\varphi}]_{\varphi_0}$ is the variance of the estimator $\hat{\varphi}$ for the physical parameter $\varphi$ evaluated at a realization $\varphi_0$. $F_\varphi(\varphi_0, T)$ is the QFI associated to the parameter $\varphi$ at measurement time $T$ evaluated at $\varphi_0$. In the following, we will consider $\delta \varphi = \sqrt{\Var[\hat{\varphi}]_{\varphi_0}}$ as a measure for the estimation error for the parameter $\varphi$. The QFI can be computed, using the symmetric logarithmic derivative (SLD) $D$ with \cite{Macieszczak2014}
\begin{align}
    \partial_\varphi \rho = \frac{1}{2} \left( D\rho + \rho D\right),
    \label{eq:symmetric logarithmic derivative}
\end{align}
where $\rho$ is the global density matrix. The QFI is given in terms of the SLD as \cite{Macieszczak2014}
\begin{align}
    F_\varphi(\varphi ,T) = \Var[D]_{\varphi},
\end{align}
meaning that it only depends on the global quantum state $\rho_\varphi$ and the form of the SLD. Since the QFI is optimized over all possible measurements, it does not depend on or specifies measurements \cite{WisemanMilburn2009}. For a pure quantum state $\rho_\varphi = \ketbra{\psi_\varphi}{\psi_\varphi}$, the QFI takes the more simple form \cite{Macieszczak2014,  Gammelmark2014}
\begin{align}
    \begin{split}
    F_\varphi(\varphi,T) &=  4\left( \braket{\psi'_{\varphi}} - |\braket{\psi'_{\varphi}}{\psi_{\varphi}}|^2\right)\\
    &=4\partial_{\varphi_1}\partial_{\varphi_2} \ln \braket{\psi_{\varphi_1}}{\psi_{\varphi_2}},
    \label{eq:qfi for pure state}
\end{split}
\end{align}
where $\ket{\psi'_\varphi}=\partial_\varphi \ket{\psi_\varphi}$ is the derivative of the state. In the case of a quantum system described by Eq. (1), the global system and environment state at time $T$ can be approximated by the matrix product state \cite{Gammelmark2014, Cabot2025}
\begin{align}
    \ket{\psi_\varphi} = \sum_{i_1 , \dots , i_n=0}^1 K_{i_n}\cdots K_{i_1}\ket{\chi_\mathrm{S}}\otimes \ket{i_1,\dots,i_n},
    \label{eq:mps}
\end{align}
where we have discretized time into $n$ steps of size $\delta t$, and introduced the Kraus operators $K_0 = 1-i\delta t \hat{H} - \kappa \delta t \hat{L}^\dagger \hat{L} /2$ and $K_1 = \sqrt{\kappa \delta t}\hat{L} e^{-i\varphi}$. The choice of the Kraus operators given here assumes an ideal photoncounting picture of the emission statistics. A similar description can be found, assuming ideal homodyne detection \cite{Mattes2025}. In the limit of infinitesimal time steps, the MPS \eqref{eq:mps} becomes a continuous matrix product state (cMPS) and one recovers the dynamics of \eqref{eq:btc lme} \cite{Macieszczak2014, Verstraete2010, Haegeman2013}. $\ket{\chi_\mathrm{S}}$ is the inital system state and $\ket{i_1, \dots ,i_n}$ captures the time record of detected emissions into a bosonic environment. 

Using the MPS description and the second equation of \eqref{eq:qfi for pure state}, the QFI rate in the long-time limit reads
\begin{align}
    f_\varphi:= \lim_{T \to \infty}\frac{F_\varphi(\varphi,T)}{T} = 4 \partial_{\varphi_1} \partial_{\varphi_2} \lambda_0 (\Delta\varphi)\big|_{\Delta\varphi=0},
    \label{eq:qfi dominant eigval}
\end{align}
where $\lambda_0 (\Delta \varphi)$ is the eigenvalue with largest real part of the deformed master equation
\begin{align}
    \partial_t \rho = \mathcal{L}\left[ \rho\right] + \kappa \left( e^{-i\Delta\varphi} - 1\right) \hat{L} \rho \hat{L}^\dagger,
    \label{eq:deformed_me_qfi}
\end{align}
with $\mathcal{L}$ the Lindbladian describing evolution according to Eq. (1). 

Alternatively, we determine the QFI by computing the inner products in the first line of \eqref{eq:qfi for pure state} \cite{Gammelmark2014}. This yields
\begin{align}
\begin{split}
    \braket{\psi_\varphi}{\psi'_\varphi} &= \sum_{j=1}^n \Tr \left[ \rho (t_{j-1}) (-i\kappa\delta t) \hat{L}^\dagger \hat{L} \right]\\
    &\overset{\delta t \to 0}{\longrightarrow}-i \kappa \int_0^T dt \Tr \left[ \rho (t) \hat{L}^\dagger \hat{L} \right] \\
    \braket{\psi'_\varphi}{\psi'_\varphi} &= \sum_{j=1}^n \Tr \left[ \rho (t_{j-1}) \kappa \delta t \hat{L}^\dagger \hat{L} \right] \\
    &+ 2\sum_{j>k = 1}^n \Tr \left[ \kappa^2 \delta t^2 \hat{L}^\dagger \hat{L} (\hat{L} \rho (t_{k-1}) \hat{L}^\dagger)(\tau) \right]\\
    &\overset{\delta t \to 0}{\longrightarrow} \int_0^T dt \Tr \left[ \rho (t) \hat{L}^\dagger \hat{L} \right] + \\
    &2\kappa^2 \int_0^T dt\int_0^{T-t} d\tau \underset{=:C(\tau)}{\underbrace{\Tr \left[ \hat{L}^\dagger \hat{L} e^{\mathcal{L}\tau}(\hat{L} \rho (t) \hat{L}^\dagger) \right]}},
\end{split}
\end{align}
with $t_k = k\delta t$, and we defined $\tau := t_{j-1} - t_{k+1}$. We further adopt the convention for the time evolution of an operator $\hat{\mathcal{O}}(t) = e^{\mathcal{L}t}\hat{\mathcal{O}}(0)$. In the long-time limit, we assume the replacement $\rho(t) \overset{t\to\infty}{\longrightarrow} \rho_\mathrm{ss}$, which yields the expression Eq. (3) for the QFI rate and shows that the QFI is related to the first and second intensity correlation functions \cite{WisemanMilburn2009} and independent of the true value of the parameter $\varphi_0$. Results for the BTC are then obtained by using $\hat{H} = \omega\hat{S}_\mathrm{x}$ and $\hat{L} = \hat{S}_-$. In the following, we will assume this choice of the Hamiltonian and jump operator if not stated otherwise, yielding the Lindblad master equation 
\begin{align}
    \partial_t  \rho = -i\omega[\hat{S}_\mathrm{x}, \rho] + \kappa \mathcal{D}[\hat{S}_-]\rho.
    \label{eq:btc lme}
\end{align}

\begin{figure}[b]
    \label{fig:deformed_me_qfi_num_res}
    \centering
    \vspace{-0.3cm}
    \includegraphics{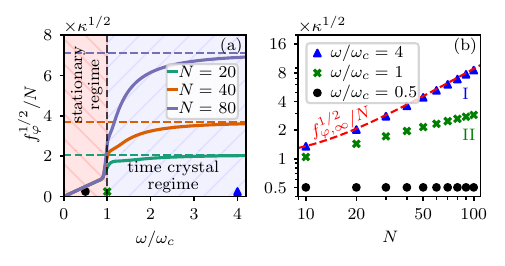}
    \vspace{-0.5cm}
    \mycaption{Scaling of the QFI rate with N.}{(a) Square root of the system size rescaled long-time limit QFI rate as a function of the Rabi frequency $\omega$ and for varying system sizes $N$. The dashed lines indicate the saturating value $f_{\varphi,\infty}$. (b) Square root of the system size rescaled QFI rate as a function of the system size $N$ for selected constant values of $\omega/\omega_\mathrm{c}$. For \MakeUppercase{\romannumeral 1} and \MakeUppercase{\romannumeral 2}, the asymptotic behavior is determined by a power law fit yielding scaling behaviors of $f_\varphi \propto N^{\alpha_i}$ with $\alpha_{\text{\MakeUppercase{\romannumeral 1}}} = 3.923 \pm 0.004$ and $\alpha_{\text{\MakeUppercase{\romannumeral 2}}} = 2.8417 \pm 0.0018$. The red dashed line indicates the saturating value $f_{\varphi,\infty}$. For both panels, numerical values are obtained, using diagonalization of \eqref{eq:deformed_me_qfi} and $f_{\varphi,\infty}$ corresponds to Eq. (4).}
    \vspace{-0.5cm}
\end{figure}


\section{Time-correlations in the single system and the superspin method}\label{SecSM:superspin}

\begin{figure*}[t]
    \centering
    \vspace{-0.7cm}
    \includegraphics{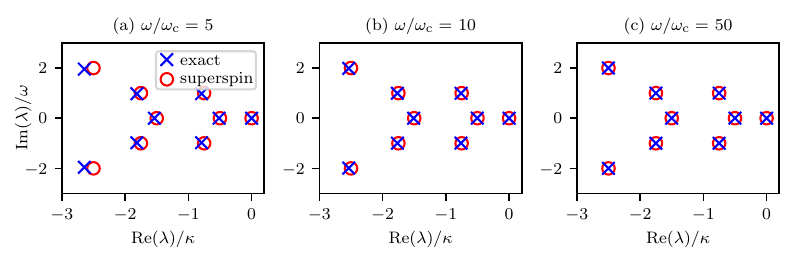}
    \vspace{-0.5cm}
    \mycaption{Benchmark of the superspin eigenvalue spectrum.}{Comparison of the Lindbladian eigenvalues for the BTC obtained by exact diagonalization (blue crosses) and by the superspin method (red circles), for different values of the Rabi frequency $\omega$ and system size $N=10$. With increasing Rabi frequency the agreement between the approximation and the exact eigenvalues increases.}
    \vspace{-0.5cm}
    \label{fig:eigenvalue spectrum superspin}
\end{figure*}

\begin{figure*}[t]
    \centering
    \includegraphics{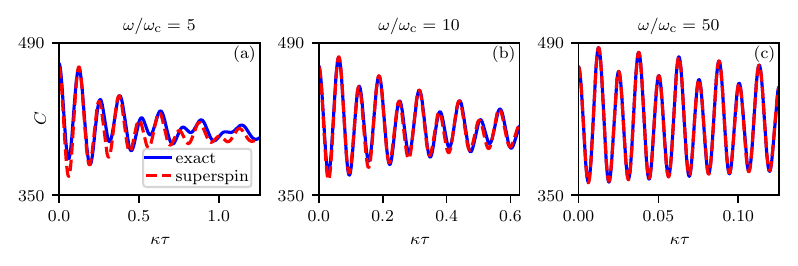}
    \vspace{-0.5cm}
    \mycaption{Benchmark of the superspin correlation function.}{Correlation function $C(\tau)$ as a function of time, for different Rabi frequencies $\omega$. In all panels, the system size is $N=10$. The exact results (blue solid line) are obtained, using numerical integration of the Lindblad master equation and the superspin results (red dashed line) represent the evaluation of \eqref{eq:superspin correlation approximation}. With increasing Rabi frequency the agreement between the superspin method and exact evolution increases.}
    \vspace{-0.5cm}
    \label{fig:numerical benchmark superspin approximation}
\end{figure*}

In this section, we compute an approximation to the correlation function $C (\tau)$ in the extreme time crystal limit $\omega/\omega_\mathrm{c} \to \infty$ of a BTC, using the superspin method developed in \cite{Nemeth2025}. This method uses the vectorized form of the Lindbladian superoperator and the density matrix. This approach assumes the transformation of operators to vectors according to $\mathcal{O} = \ketbra{m}{n} \to \superket{\mathcal{O}} = \ket{m} \otimes \ket{n}^{*}$. Applying this transformation to the Lindblad Master Equation \eqref{eq:btc lme} leads to the matrix equation $\partial_t \superket{\rho} = \Tilde{\mathcal{L}}\superket{\rho}$, with the vectorized Lindblad superoperator \cite{Nemeth2025}
\begin{align}
    \Tilde{\mathcal{L}} = \Tilde{\mathcal{L}}_0 + \Tilde{\mathcal{L}}_\mathrm{D} = \underset{\Tilde{\mathcal{L}}_0}{\underbrace{i\omega\hat{J}_\mathrm{x}}} -\underset{\Tilde{\mathcal{L}}_\mathrm{D}}{\underbrace{\frac{\kappa}{4}\left( \hat{J}_\mathrm{x}^2 + \hat{\Vec{J}}^2 \right)}},
\end{align}
where we have defined the superspin components $\hat{J}_\alpha = \hat{S}_\alpha \otimes \mathds{1} - \mathds{1} \otimes \hat{S}_\alpha$. Since the vectorized Lindblad superoperator only depends on the total superspin and its $\mathrm{x}-$projection, it is customary to use the $\mathrm{x}-$superspin basis of the respective Liouville space, where the common eigensuperkets can be labeled as $\superket{j,j_\mathrm{x}}$ \cite{Nemeth2025}. We restrict the calculations to the maximally polarized sector with a BTC composed of $N$ emitters. This implies values of $j=0,\dots,N$ and $j_\mathrm{x}=-j,\dots,j$ \cite{Nemeth2025}.

In the strong driving or extreme time crystal limit ($\omega/\omega_\mathrm{c}\to \infty$) and for fixed system size, the term $\Tilde{\mathcal{L}}_\mathrm{D}$ can be considered a small perturbartion of the unperturbed contribution $\Tilde{\mathcal{L}}_0$. Using a first order perturbation calculation, we compute an approximation to the Lindbladian eigenvalues as \cite{Nemeth2025}
\begin{align}
    \lambda_{j,j_\mathrm{x}} = i\omega j_\mathrm{x} - \frac{\kappa}{4}(j_\mathrm{x}^2 + j(j+1)),
\end{align}
where the labels $j,j_\mathrm{x}$ correspond to the $\mathrm{x}-$basis superspin states. We are interested in computing the correlation function $C(\tau)$. In the extreme time crystal limit, the stationary state approaches the maximally mixed state, i.e. $\rho_\mathrm{ss} \approx \mathds{1}/(N+1)$ \cite{Puri1979, Hannukainen2018}. Thus, to first order the correlation function reads \cite{Albarelli2025}
\begin{align}
    C(\tau) \approx \sum_{j,j_\mathrm{x}} e^{\lambda_{j,j_\mathrm{x}}\tau} \langle\langle \hat{S}_+ \hat{S}_-| j,j_\mathrm{x} \rangle \rangle\langle\langle j,j_\mathrm{x} | \hat{S}_- \hat{S}_+ \rangle \rangle /(N+1),
\end{align}
where we used the decomposition $e^{\mathcal{L}\tau} \approx \sum_{j,j_\mathrm{x}} e^{\lambda_{j,j_\mathrm{x}}\tau} \superket{j,j_\mathrm{x}}\superbra{j,j_\mathrm{x}}$. We further simplify this expression by determining the $\mathrm{x}-$basis superspin states, using the rule $\hat{J}_\alpha \superket{\hat{O}} = \superket{[\hat{S}_\alpha , \hat{O}]}$ \cite{Nemeth2025, Albarelli2025}. For example the state corresponding to $j=1, j_\mathrm{x}=0$ reads up to a normalization and phase
\begin{align}
    \hat{J}_\mathrm{x} \superket{\hat{S}_\mathrm{x}} &= \superket{[\hat{S}_\mathrm{x},\hat{S}_\mathrm{x}]} = 0  \\
    \hat{J}^2 \superket{\hat{S}_\mathrm{x}} &=\superket{[\hat{S}_\mathrm{y},[\hat{S}_\mathrm{y},\hat{S}_\mathrm{x}]]} + \superket{[\hat{S}_\mathrm{z},[\hat{S}_\mathrm{z},\hat{S}_\mathrm{x}]]}= 2 \superket{\hat{S}_\mathrm{x}}.
\end{align}
We thus conclude $\superket{1,0} = \mathcal{N}_{1,0} \superket{\hat{S}_\mathrm{x}}$ and determine the normalization up to a phase
\begin{align}
    |\mathcal{N}_{1,0}|^2 = \frac{1}{\Tr \small[ \hat S_\mathrm{x}^2\small]} = \frac{12}{N(N+1)(N+2)}.
\end{align}
Next, we decompose the superstates $\superket{\hat{S}_\pm \hat{S}_\mp}$ in the $\superket{j,j_\mathrm{x}}$ basis by using the ladder operators in the $\mathrm{x}-$basis $\hat{S}_\pm^{(\mathrm{x})} = \hat{S}_\mathrm{y}\pm i \hat{S}_\mathrm{z}$ and applying the identity for ladder operators in the $\mathrm{z}$-basis $\hat{S}_\pm \hat{S}_\mp = \hat{\Vec{S}}^2 - \hat{S}_\mathrm{z}^2 \pm \hat{S}_\mathrm{z}$, yielding
\begin{widetext}
\begin{align}
    \superket{\hat{S}_\pm\hat{S}_\mp} &= \frac{2}{3} \underset{\propto \superket{0,0}}{\underbrace{\superket{\hat{\Vec{S}}^2}}} \mp \frac{i}{2}  \underset{\propto (\superket{1,1}-\superket{1,-1})}{\underbrace{\left(\superket{\hat{S}_+^{(\mathrm{x})}} - \superket{\hat{S}_-^{(\mathrm{x})}} \right)}} + \frac{1}{2} \underset{\propto\superket{2,0}}{\underbrace{\left( \superket{\hat{S}_\mathrm{x}^2} - \frac{1}{3} \superket{\hat{\Vec{S}}^2} \right)}} + \frac{1}{4}\underset{\propto(\superket{2,2} + \superket{2,-2})}{\underbrace{\left( \superket{(\hat{S}_+^{(\mathrm{x})})^2} + \superket{(\hat{S}_-^{(\mathrm{x})})^2} \right)}},
    \label{eq:spmsmp in superbasis}\\
    \superket{0,0} &= \frac{4}{\sqrt{N+1}N(N+2)}\superket{\hat{\Vec{S}}^2}, \quad \superket{1,\pm1} = \sqrt{\frac{6}{N(N+1)(N+2)}} \superket{\hat{S}_\pm^{(\mathrm{x})}},\label{eq:basis states start}\\
    \quad\superket{2,0} &= \sqrt{\frac{180}{(N-1)N(N+1)(N+2)(N+3)}} \left(\superket{\hat{S}_\mathrm{x}^2} - \frac{1}{3}\superket{\hat{S}^2}\right),\\
    \superket{2,\pm2} &= \sqrt{\frac{30}{(N-1)N(N+1)(N+2)(N+3)}}  \superket{(\hat{S}_\pm^{(\mathrm{x})})^2}. \label{eq:basis states end}
\end{align}
\end{widetext}
where the basis states underset in \eqref{eq:spmsmp in superbasis} are given by Eqs. \eqref{eq:basis states start} - \eqref{eq:basis states end}. We chose the normalization constants to be real. With this decomposition and the respective eigenvalues $\lambda_{j,j_\mathrm{x}}$, we compute the correlation function as
\begin{align}
\begin{split}
    C(\tau) &\approx  N(N+2)\Biggr[\frac{N(N+2)}{36} - \frac{e^{-3\kappa\tau/4} \cos(\omega\tau)}{12}  \\
    &+ \frac{N^2 +2N -3}{240} \left( e^{-5\kappa\tau/2}\cos(2\omega\tau) + \frac{e^{-3\kappa\tau/2}}{3} \right)\Biggr].
    \label{eq:superspin correlation approximation}
\end{split}
\end{align}
With this expression, we compute the QFI, which yields the result in Eq. (4) in the extreme time crystal limit $\omega/\omega_\mathrm{c}\to \infty$. We numerically benchmark the expressions for the first order approximated eigenvalues in \figref{fig:eigenvalue spectrum superspin} and the correlation function in \figref{fig:numerical benchmark superspin approximation}.


\section{Large deviations approach and estimation error}\label{SecSM:largedeviations}

\subsection{Photoncounting}

In the main text, we consider estimation of the parameter $\varphi$ via continuous monitoring of the decoder output intensity $I_T = (1/T)\int_0^TdN(t)$. The statistics of the number of photon emissions until time $T$, $M = \int_0^T dN(t)$, are described by a Laplace transform of the probability distribution, yielding the generating function \cite{Garrahan2010}
\begin{align}
    Z_\mathrm{c}(s, \Delta\varphi) :=\sum_{M=0}^\infty p_M(T,\Delta\varphi)e^{-sM},
\end{align}
with the probability $p_M(T, \Delta\varphi)$ of observing $M$ emissions until time $T$. The moments of the photon count at time $T$ are then given by derivatives of $Z_\mathrm{c}$. In the long-time limit, the generating function has the large deviation form $Z_\mathrm{c}(s) \sim e^{T\theta_\mathrm{c}(s,\Delta\varphi)}$ with the large deviation function $\theta_\mathrm{c}(s,\Delta\varphi)$ \cite{Garrahan2010}. $Z_\mathrm{c}$, $p_M$ and $\theta_\mathrm{c}$ inherit a $\Delta\varphi$-dependence from the cascaded Master equation, since the operators $\hat{L}_\mathrm{casc}$ and the stationary state $\rho_\mathrm{ss}$ depend on $\Delta\varphi$ in the cascaded system. $\theta_\mathrm{c}(s,\Delta\varphi)$ is computed as the dominant eigenvalue of the tilted master equation \cite{Zoller1987, Carollo2018, Garrahan2010, Cabot2024}
\begin{align}
    \partial_t \rho = \mathcal{L}_\text{casc} [\rho]+ (e^{-s} -1) \hat{L}_\text{casc}\rho \hat{L}_\text{casc}^\dagger,
    \label{eq:tilted me cascaded system}
\end{align}
with the physical Lindbladian $\mathcal{L}_\text{casc}$ of the cascaded system and a bias described by $s$. For $s=0$, \eqref{eq:tilted me cascaded system} generates the physical dynamics and for $s\neq0$ it generates an ensemble of trajecotries with the bias $e^{-sM}$ \cite{Garrahan2010, Zoller1987}. The estimation error thus is approximated in terms of the large deviations function $\theta_\mathrm{c}(s,\Delta\varphi)$ as
\begin{align}
    \delta \varphi \sim \frac{\overline{\delta\varphi}}{\sqrt{T}} \text{, with }\overline{\delta\varphi} :=  \frac{\sqrt{\partial_s^2 \theta_\mathrm{c}(s,\Delta\varphi)}}{|\partial_{\varphi}\partial_s \theta_\mathrm{c}(s,\Delta\varphi)|}\biggr|_{s=0} = \frac{\overline{\sigma_{I_T}}}{|\partial_\varphi I_T|} ,
\end{align}
with the time rescaled standard deviation $\overline{\sigma_{I_T}} = \sqrt{\partial_s^2 \theta_\mathrm{c}(s,\Delta\varphi)}|_{s=0}$, and the derivative of the long-time averaged intensity $-\partial_\varphi \partial_s \theta_\mathrm{c}(s,\Delta\varphi)|_{s=0} = \partial_\varphi I_T$. It follows immediately that the estimation error decreases with measurement time as $\delta\varphi\sim 1/\sqrt{T}$, obeying the standard quantum limit for long measurement times. 

\subsection{Homodyne detection}

In the main text, we consider the time averaged homodyne current $J_T = (1/T)\int_0^T J_\beta (t) dt = K_T/T$, with the time integrated homodyne current $K_T$. Its moments are determined by the generating function
\begin{align}
    Z_\mathrm{h}(s, \varphi-\beta) = \int dK e^{-sK}p_K(T,\varphi-\beta),
\end{align}
where $p_K(T,\varphi-\beta) = P(K_T = K)$ is the probability density of observing $K_T=K$. For long times $T$, $Z_\mathrm{h}$ acquires the large deviation form $Z_\mathrm{h}(s)\sim e^{T\theta_\mathrm{h}(s,\varphi-\beta)}$ \cite{Cabot2023}, with the large deviation function $\theta_\mathrm{h}(s,\varphi-\beta)$. $Z_\mathrm{h}$, $p_K$ and $\theta_\mathrm{h}$ inherit a $\varphi-\beta$ dependence from $J_\beta$. $\theta_\mathrm{h}$ is computed as the eigenvalue with largest real part of the deformed master equation \cite{Cabot2023}
\begin{align}
    \partial_t \rho = \mathcal{L}[\rho] - s\sqrt{\kappa}\left(\hat{S}_- e^{i(\beta-\varphi)}\rho + \rho \hat{S}_+e^{-i(\beta-\varphi)} \right) + \frac{s^2}{2}\rho,
    \label{eq:deformed me homodyne}
\end{align}
with the physical Lindbladian of a single BTC $\mathcal{L}$. For $s\to0$, the physical evolution is recovered. $J_T$ approaches its ensemble average in the long-time limit, as in a law of large numbers, yielding $\lim_{T\to\infty}J_T = \lim_{T\to\infty} \mathbb{E}[J_T] = \lim_{T\to\infty}\mathbb{E}[K_T]/T=  -\partial_s \theta_\mathrm{h}(s,\varphi-\beta)$ \cite{WisemanMilburn2009}, and its variance behaves as $\lim_{T\to\infty} \mathbb{E}[J_T^2]-\mathbb{E}[J_T]^2 = \lim_{T\to\infty}\partial_s^2 \theta_\mathrm{h}(s,\varphi-\beta)/T$. The estimation error for measuring $\varphi$ via monitoring of the long-time averaged homodyne current thus is approximated by
\begin{align}
    \delta\varphi\sim\frac{\overline{\delta\varphi}}{\sqrt{T}}, \text{ with } \overline{\delta\varphi} := \frac{\sqrt{\partial_s^2 \theta_\mathrm{h}(s,\varphi-\beta)}}{|\partial_\varphi \partial_s \theta_\mathrm{h}(s,\varphi-\beta)|}\biggr|_{s=0} = \frac{\overline{\sigma_{J_T}}}{|\partial_\varphi J_T|},
\end{align}
with the time rescaled standard deviation $\overline{\sigma_{J_T}}:= \sqrt{\partial_s^2 \theta_\mathrm{h}(s,\varphi-\beta)}|_{s=0}$, the time rescaled estimation error $\overline{\delta\varphi}$ and the derivative of the homodyne current in the long-time limit $\partial_\varphi J_T := -\partial_\varphi \partial_s \theta_\mathrm{h}(s,\varphi-\beta)|_{s=0}$. Similar to the estimation error in the cascaded measurement protocol, the estimation error decreases with measurement time as $\delta\varphi\sim 1/\sqrt{T}$, obeying the standard quantum limit for long measurement times. 


\section{Holstein Primakoff approach to the single system}
\label{SecSM:HP approach single system}

In the stationary regime, we approximate the behavior of the BTC by applying the Holstein-Primakoff (HP) transformation to its spin operators. Here we follow the prescription given in \cite{Cabot2024} and express the spin operators in terms of bosonic ladder operators
\begin{align}
    \hat{S}_+ = \hat{b}^{\dagger} \sqrt{2S - \hat{b}^{\dagger}\hat{b}}, \hspace{0.2cm} \hat{S}_-=\sqrt{2S - \hat{b}^{\dagger}\hat{b}}\hat{b},
\end{align}
with $S=N/2$ as in the main text. The bosonic mode described by the operators $\hat{b}^{(\dagger)}$ is assumed to be in a large displaced state, such that only quantum fluctuations around this large displaced state are studied
\begin{align}
    \hat{b} \to \hat{b}+ \sqrt{S}\beta,
\end{align}
where $\beta$ is a complex field, which is determined below. If $S$ describes a large displacement, the spin operators can be expanded in the small parameter $\epsilon = 1/\sqrt{S}$
\begin{align}
    \hat{m}_{\alpha} = \frac{\hat{S}_{\alpha}}{S} = \sum_{l=0}^{\infty} \epsilon^l \hat{m}_{\alpha, l}.
    \label{eq:rescaled operators single system}
\end{align}
Explicit expressions up to order $\mathcal{O}(\epsilon)$ are given by
\begin{align}
    m_{+,0} = \sqrt{k}\beta^{*} \text{ and } \hat{m}_{+,1} = \frac{1}{2\sqrt{k}}[(4-3|\beta|^2)\hat{b}^{\dagger} - \beta^{*2}\hat{b}],
\end{align}
where $k=2-|\beta|^2$. The field $\beta$ is then determined by inserting the expanded operators in the Lindblad master equation. However, it is helpful to account for the physical behaviour of the system before. The critical frequency scales with the system size as $\omega_c = N\kappa/2 = S\kappa$. This extensive character of the Rabi frequency is accounted for by rescaling according to
\begin{align}
    \omega = \Tilde{\omega}S, \hspace{0.2cm} \tau = St,
    \label{eq:rescaled parameters}
\end{align}
where also the time scale of the dynamics is rescaled. With the rescaled operators and parameters, the Lindblad master equation is obtained
\begin{align}
    \partial_{\tau} \rho= -i\Tilde{\omega}S[\hat{m}_\mathrm{x}, \rho] + \kappa S \left(\hat{m}_- \rho \hat{m}_+ - \frac{1}{2}\{\hat{m}_+ \hat{m}_-, \rho\}\right).
    \label{rescaled-ME}
\end{align}
By expanding the density matrix $\rho = \sum_{l=0}^{\infty} \epsilon^l \rho_l$ and setting the right-hand side of \eqref{rescaled-ME} equal to zero, we obtain the field $\beta$ in the stationary state. Up to $\mathcal{O}(\epsilon)$, this yields
\begin{align}
    0=-i\Tilde{\omega} \pm \kappa m_{\pm,0} \Rightarrow \beta \sqrt{2-|\beta|^2} = -i\frac{\Tilde{\omega}}{\kappa},
\end{align}
from which $\beta$ is determined to $\beta=-i\sqrt{1-\sqrt{1-(\Tilde{\omega}/\kappa)^2}}$. This is only valid in the regime where $\Tilde{\omega} \leq \kappa$, i.e. the stationary regime. For the dominant order of the rescaled ladder operators, we obtain
\begin{align}
    m_{\pm, 0} = \pm \frac{i \Tilde{\omega}}{\kappa}.
    \label{eq:zero order ladder operators single system}
\end{align}
The term of $\mathcal{O}(1)$ of the rescaled Lindblad master equation yields
\begin{align}
    \partial_{\tau}\rho_0 = \kappa \left(\hat{m}_{-,1} \rho_0 \hat{m}_{+,1} - \frac{1}{2}\{\hat{m}_{+,1}\hat{m}_{-,1}, \rho_0\}\right),
\end{align}
which we solve for the stationary state by the ansatz $\rho_{0,\text{ss}} = \ketbra{E_0}{E_0}$, where the state $\ket{E_0}$ is annihilated by the fluctuation of the lowering operator $\hat{m}_{-,1}$. By expanding $\hat{m}_{-,1} = A\hat{b}+B\hat{b}^{\dagger}$ with the coefficients
\begin{align}
    A = \frac{1+3\sqrt{1-(\Tilde{\omega}/\kappa)^2}}{2\sqrt{1+\sqrt{1-(\Tilde{\omega}/\kappa)^2}}}, \hspace{0.1cm} B = \frac{1-\sqrt{1-(\Tilde{\omega}/\kappa)^2}}{2\sqrt{1+\sqrt{1-(\Tilde{\omega}/\kappa)^2}}},
\end{align}
we arrive at the stationary state
\begin{align}
    \ket{E_0} = \frac{1}{\sqrt{\mathcal{N}}}\ket{0} + \frac{1}{\sqrt{\mathcal{N}}}\sum_{n=1}^{\infty}(-1)^n \left( \frac{B}{A}\right)^n \sqrt{\frac{(2n-1)!!}{2n!!}} \ket{2n}.
    \label{eq:hp approx stationary state}
\end{align}
$\mathcal{N}$ is a normalization constant and $\ket{n}$ is a Fock state in the bosonic mode. Note that this is only well defined for $|B/A| \leq 1$ and in the stationary regime $\Tilde{\omega}<\kappa$, meaning that the given HP approach is only valid in the stationary regime.

\subsection{Calculation of the QFI}

\begin{figure}[t]
    \centering
    \vspace{-0.3cm}
    \includegraphics{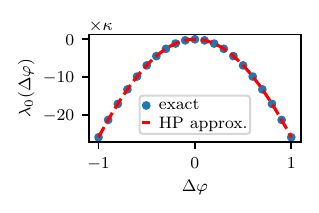}
    \vspace{-0.5cm}
    \mycaption{Benchmark of the HP approximation for the QFI.}{Dominant eigenvalue of \eqref{eq:deformed_me_qfi} as a function of the phase difference $\Delta\varphi = \varphi_1-\varphi_2$. Comparison of the dominant eigenvalue $\lambda_0$ computed via exact diagonalization \eqref{eq:deformed_me_qfi} (blue dots) and the HP approximation \eqref{eq:hp eigvals qfi} (red dashed line), for $N=30$, and in the stationary regime $\omega/\omega_{c} = 0.5$.}
    \label{fig:hp approximation eigenvalue qfi}
    \vspace{-0.25cm}
\end{figure}

In order to determine an expression for the QFI, we analyze the deformed master equation \eqref{eq:deformed_me_qfi}. Inserting the rescaled operators and parameters yields the rescaled deformed master equation
\begin{align}
    \partial_{\tau} \rho = \mathcal{L}[\rho] + \left( e^{-i \Delta \varphi } -1 \right) \kappa S \hat{m}_- \rho \hat{m}_+.
\end{align}
We make an ansatz for the dominant eigenstate similar to the undeformed master equation as
\begin{align}
    \rho_0 = \ketbra{E_0^{\varphi_1}}{E_0^{\varphi_2}}.  
\end{align}
$\ket{E_0^\varphi}$ denotes the stationary state of the fluctuations $\hat{m}_{-,1}$ as in \eqref{eq:hp approx stationary state}, evaluated at the point $\varphi$. As this state is independent of $\varphi$, one could neglect the superscript, but it is carried here in order to keep the notation consistent with situations, where the stationary state depends on the estimated parameter \cite{Cabot2024}. We expand the action of $\hat{m}_-$ as
\begin{align}
    \hat{m}_- \ket{E_0^{\varphi_1}} &= \left( m_{-,0}^{\varphi_1} + \frac{1}{\sqrt{S}}\hat{m}_{-,1}^{\varphi_1} + ... \right) \ket{E_0^{\varphi_1}}, \\
    \bra{E_0^{\varphi_2}} \hat{m}_- &= \bra{E_0^{\varphi_2}} \left( m_{-,0}^{\varphi_2} + \frac{1}{\sqrt{S}}\hat{m}_{-,1}^{\varphi_2} + ... \right).
\end{align}
With this ansatz, the deformed master equation yields
\begin{widetext}
\begin{align}
    \partial_{\tau} \rho_0= \mathcal{L} [\rho_0] + \left( e^{-i \Delta \varphi} -1\right) \kappa S \left( m_{-,0}^{\varphi_1} + \frac{1}{\sqrt{S}}\hat{m}_{-,1}^{\varphi_1} + ... \right) \ketbra{E_0^{\varphi_1}}{E_0^{\varphi_2}} \left( m_{-,0}^{\varphi_2} + \frac{1}{\sqrt{S}}\hat{m}_{-,1}^{\varphi_2} + ... \right),
\end{align}
\end{widetext}
with $\Delta\varphi:=\varphi_1 - \varphi_2$ and $m_{\pm,0}^\varphi = m_{\pm,0} = \pm i \Tilde{\omega}/\kappa$ as before. Of interest are the derivatives of the dominant eigenvalue of this tilted master equation, $\lambda_0(\Delta\varphi)$. The contribution $\mathcal{L}[\rho_0]$ vanishes due to $\hat{m}_{-,1}^{\varphi}\ket{E_0^{\varphi}} = 0$. The leading contribution to $\lambda_0$ reads
\begin{align}
\begin{split}
    &\left( e^{-i \Delta \varphi} -1\right) \kappa S m_{-,0}^{\varphi_1} m_{+,0}^{\varphi_2} \ketbra{E_0^{\varphi_1}}{E_0^{\varphi_2}} \\
    \quad&= \left( e^{-i \Delta \varphi} -1\right) \kappa S \frac{\Tilde{\omega}^2}{\kappa^2}\rho_0 \\
    &= \underset{=: \Tilde{\lambda}_0 (\varphi_1, \varphi_2)}{\underbrace{\left( e^{-i \Delta \varphi} -1\right) S \frac{\Tilde{\omega}^2}{\kappa}}}\rho_0 .
\end{split}
\end{align}
The next-to-dominant term vanishes, because $\hat{m}_{-,1}^{\varphi}\ket{E_0^{\varphi}} = 0$. Thus, one finds for the dominant eigenvalue of the deformed master equation in leading order, $\lambda_0 (\Delta\varphi) = S \Tilde{\lambda}_0 (\Delta\varphi )$, where the factor of $S$ comes from the previous rescalings \eqref{eq:rescaled parameters}. This expression is benchmarked with results from diagonalization of \eqref{eq:deformed_me_qfi} in \figref{fig:hp approximation eigenvalue qfi}. In the stationary regime, the QFI rate in the long-time limit thus has the compact form
\begin{align}
\begin{split}
    f_\varphi &= \lim_{T \to \infty} \frac{F_\varphi (\varphi, T)}{T} = 4 \partial_{\varphi_1} \partial_{\varphi_2}  \lambda_0 \big|_{\varphi_1 = \varphi_2} \\
    &= 4 \partial_{\varphi_1} \partial_{\varphi_2} \left(e^{-i (\varphi_1 - \varphi_2)} -1\right) \frac{S^2 \Tilde{\omega}^2}{\kappa}\big|_{\varphi_1 = \varphi_2}\\
    &= 4 \frac{\omega^2}{\kappa}.
    \label{eq:hp eigvals qfi}
\end{split}
\end{align}

\subsection{Estimation error for the homodyne protocol}

\begin{figure}[t]
    \centering
    \vspace{-0.3cm}
    \includegraphics[width=\linewidth]{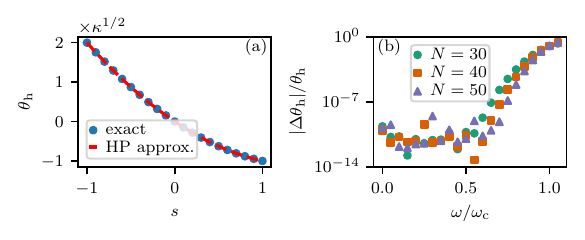}
    \vspace{-0.5cm}
    \mycaption{Benchmark of the HP approximation for the homodyne protocol.}{(a) Dominant eigenvalue of \eqref{eq:deformed me homodyne} as a function of the bias parameter $s$ for $\varphi-\beta=0.1$, fixed system size $N=30$, and in the stationary regime $\omega/\omega_\mathrm{c}=0.5$. Comparison of values obtained by exact diagonalization of  \eqref{eq:deformed me homodyne} (blue dots) and evaluation of \eqref{eq:hp approximation homodyne} (red dashed line). (b) Relative error of the dominant eigenvalue $\theta_\mathrm{h}$ as a function of the Rabi frequency $\omega$, for different system sizes, and fixed values $\varphi-\beta=0.1$ and $s=-0.05$. The error is defined as the absolute difference of the HP approximated value and the exact diagonalized value, divided by the exact value. One observes that the error increases towards the critical Rabi frequency, but is small in a large portion of the stationary regime.}
    \label{fig:hp approximation homodyne estimation}
\end{figure}

Here, we compute an approximation to the estimation error of the parameter $\varphi$ via homodyne detection, using the HP approach. With the rescaled parameters \eqref{eq:rescaled parameters} and the expanded operators \eqref{eq:rescaled operators single system}, the deformed master equation for the homodyne current \eqref{eq:deformed me homodyne} reads
\begin{align}
    \partial_{\tau} \rho = \mathcal{L}[\rho] +s\sqrt{\kappa}\left( \hat{m}_- e^{i(\beta - \varphi)} \rho + \rho \hat{m}_+ e^{-i(\beta - \varphi)}\right) + \frac{s^2}{2}\frac{\rho}{S}.
     \label{eq:rescaled me homodyne}
\end{align}
As an ansatz for the dominant eigensatate we choose the vacuum of the fluctuations $\rho_0 = \ketbra{E_0}{E_0}$. Inserting on the right hand side and using the relations \eqref{eq:zero order ladder operators single system} yields to leading order
\begin{widetext}
\begin{align}
\begin{split}
    &s\sqrt{\kappa}\left\{ e^{i(\beta  - \varphi)}(m_{-,0} + \hat{m}_{-,1} + \dots) \rho_0 + \rho_0 (m_{+,0} + \hat{m}_{+,1}+\dots)e^{-i(\varphi - \beta)} \right\} + \frac{s^2}{2} \frac{\rho_0}{S} \\&\approx s\sqrt{\kappa}\frac{i\Tilde{\omega}}{\kappa}\left( e^{-i(\beta - \varphi)} - e^{i(\beta - \varphi)} \right) \rho_0 +\frac{s^2}{2}\frac{\rho_0}{S},
\end{split}
\end{align}
\end{widetext}
where we included the term $s^2/(2S)$, since it defines the variance of the homodyne current, but neglected terms with $\hat{m}_{\pm,2}$, which are also of $\mathcal{O}(\epsilon^2)$, but only contribute to the $-\partial_s \partial_\varphi \theta_\mathrm{h}$ term in the estimation error, where they are subdominant to $\mathcal{O}(1)$ terms. Thus, the dominant eigenvalue of the rescaled deformed master equation reads
\begin{align}
    \Tilde{\theta}_\mathrm{h}( s,\varphi-\beta) = 2s\frac{\omega}{S\sqrt{\kappa}} \sin(\beta - \varphi) + \frac{s^2}{2S},
    \label{eq:hp approximation homodyne}
\end{align}
where, due to the rescaling of \eqref{eq:deformed me homodyne},  $\theta_\mathrm{h} (s, \varphi-\beta) =  S\Tilde{\theta}_\mathrm{h}( s,\varphi-\beta)$. For the estimation error, this means
\begin{align}
    \overline{\delta\varphi} = \sqrt{\frac{\partial_{s^2}\theta_\mathrm{h}(s,\varphi-\beta)}{(\partial_{\varphi}\partial_{\Tilde{s}}\theta_\mathrm{h}(\Tilde{s},\varphi-\beta))^2}}\biggr |_{\Tilde{s}=0} = \frac{\sqrt{\kappa}}{2\omega|\cos(\beta - \varphi)|}.
\end{align}
This result is numerically benchmarked with results from diagonalization of \eqref{eq:deformed me homodyne} as shown in \figref{fig:hp approximation homodyne estimation}. 

\section{Construction of the optimal decoder in the stationary regime}
\label{SecSM:Construction of the optimal decoder in the stationary regime}

This section follows the construction of an optimal decoder presented in \cite{Yang}. We show how the perfect absorber protocol emerges from this construction as the optimal measurement protocol in the stationary regime. We assume here a matrix product (MP) description of the cascaded source-decoder system, where we access the trajectory of emissions $\{i_1, \dots, i_n\}$ in small but finite time bins of size $\delta t$ over a measurement time $T=n\delta t$ by continuously monitoring the decoder output. This is formally implemented by the measurement
\begin{align}
    \Pi_{\{i_j\}}^\text{DE} = U_\text{DE}^\dagger (T) \Pi_{\{i_j\}} U_\text{DE}(T),
\end{align}
acting on the decoder and emission field state $\rho_\mathrm{D} (0) \otimes \rho_\mathrm{E} (T)$ before their interaction. $\rho_\mathrm{D} (0) = \ketbra{\chi_\mathrm{D}(0)}{\chi_\mathrm{D}(0)}$ is the decoder initial state and $\rho_\mathrm{E}(T)$ describes the emission field at time $T$, after the interaction with the source. $\Pi_{\{i_j\}} = \bigotimes_{j=1}^n \ket{i_j}\bra{i_j}$ is the projector on the observed trajectory in the basis of the source and decoder emission field time bin modes, interacting sequentially with the source and decoder \cite{Ciccarello2022}. $U_\text{DE}$ is a unitary, describing the joint evolution of the decoder and emission field. In the MP description of the decoder and emission field state, this unitary is a MP Operator (MPO). The resulting measurement $\Pi_{\{i_j\}}^\text{DE}$ thus inherits a MP structure.

The global system and emission field state is a pure MPS, $\ket{\psi_\varphi}$, with the true value $\varphi$ inscribed. Any projective measurement $\{ \Pi_k^\text{SE} \}$ of the state $\ket{\psi_\varphi}$ containing the projector $\Pi_0^\text{SE} = \ketbra{\psi_{\varphi'}}{\psi_{\varphi'}}$ is optimal \cite{Braunstein1994,Pezze2017, Liu2020, Pezze2014} in the vicinity of $\varphi'$, in the sense that it saturates the QCRB evaluated at $\varphi'$. By tuning $\varphi'$ to $\varphi' =\varphi$, the projective measurement $\{\Pi_k^\mathrm{SE}\}$ becomes optimal for the true realization of $\varphi$. For $\Pi_{\{i_j\}}^\text{DE}$ to be optimal, it has to obey the same emission field structure as $\Pi_0^\text{SE}$ for at least one specific trajectory $\{m_j\}$, yielding the condition
\begin{align}
\begin{split}
    &U_\text{DE}(\varphi',T) \left[ \Pi_0^\text{SE} \otimes\rho_D(0) \right] U^\dagger_\text{DE}(\varphi',T)\\
    \quad&= \Pi_{\{m_j\}} \otimes \rho_\text{SD}(\varphi',T).
    \label{eq:optimal decoder}
\end{split}
\end{align}
Physically, the unitary $U_{\text{DE}}(\varphi', T)$ transforms the source and emission field state $\Pi_0^\text{SE}$ into the specific trajectory $\{m_j\}$ of the emission field as described by the projector $\Pi_{\{m_j\}}$, for $\varphi=\varphi'$. In this transformation the source and decoder evolve into a general state $\rho_\text{SD}(\varphi',T)$, which is discarded. $U_\text{DE}$ and $\rho_\text{SD}$ inherit a dependence on $\varphi'$ from $\Pi_0^\mathrm{SE}$. A convenient choice for the target trajectory is the vacuum $\{m_j\} = \{0,0,\dots,0\}$, which describes a dark decoder output and renders the decoder a perfect absorber.

A general construction of a decoder satisfying \eqref{eq:optimal decoder} can be challenging. However, in the long-time limit, a convenient construction can be found, if the source evolves according to a time-homogeneous master equation with support for a stationary state $\rho_\text{S,ss} = \sum_{k=1}^{d} p_{k,\text{ss}}(\varphi') \ket{k_\text{ss}(\varphi')}_\mathrm{S} \prescript{}{\mathrm{S}}{\bra{k_\text{ss}(\varphi')}}$ at a parameter value $\varphi'$. In the stationary regime ($\omega<\omega_\mathrm{c}$), the source stationary state is well approximated by the pure stationary state $\rho_\mathrm{S,ss} = \ket{E_0}_\mathrm{S} \prescript{}{\mathrm{S}}{\bra{E_0}}$ [see \eqref{eq:hp approx stationary state}]. In such a situation, the prescription for the decoder operators given in \cite{Yang} simplifies to $\hat{H}_\text{D} =-W_0 \hat{H}_\mathrm{S}(\varphi')W_0^\dagger$ and $\hat{L}_\text{D} =-W_0 \hat{L}_\text{S}(\varphi') W_0^\dagger$, in the sense that their matrix elements with respect to the source stationary state $\ket{E_0}_\mathrm{S}$ and decoder stationary state $\ket{E_0}_\mathrm{D} = W_0 \ket{E_0}_\mathrm{S}$ are equal. The decoder is assumed to have the same Hilbert space dimension as the source, and $W_0$ is an arbitrary unitary. The stationary state of the source-decoder system then reads $\ket{\psi} = \ket{E_0}_\mathrm{S} \otimes \ket{E_0}_\mathrm{D}$. At $\varphi=\varphi'$, this state is a pure dark state, as is shown in \cite{Cabot2024}. With the choice
\begin{align}
    W_0 = \exp [i\pi\hat{S}_\mathrm{z}^\mathrm{S}],
\end{align} 
the decoder operators are $\hat{H}_\mathrm{D} = \omega \hat{S}_\mathrm{x}^\mathrm{D}$ and $\hat{L}_\mathrm{D}= \kappa e^{-i\varphi'}\hat{S}_-^\mathrm{D}$, with $\hat{S}_\alpha^\mathrm{D}$ acting on the decoder Hilbert space and $\hat{S}_\alpha^\mathrm{S}$ on the source Hilbert space. This means that the optimal decoder is an identical copy of the source with a phase shift $\varphi'$ imprinted on emitted photons.


\section{Holstein Primakoff approach to the source-decoder system}\label{SecSM:HP absorber}
In this section, the HP approach is taken to the cascaded source-decoder system, to obtain an analytical expression for the estimation error $\delta \varphi$ in the cascaded stationary regime ($\omega <\omega_\mathrm{c,casc}$) similar to \cite{Cabot2024}. We start as before with the HP transformation
\begin{align}
    \hat{S}_+^{i} = \hat{b}_i^{\dagger} \sqrt{2S - \hat{b}_i^{\dagger}\hat{b}_i}, \hspace{0.1cm} \hat{S}_-^{i} = \sqrt{2S - \hat{b}_i^{\dagger}\hat{b}_i}\hat{b}_i,
\end{align}
where the index $i$ indicates the operators of the source ($i=\mathrm{S}$) and of the decoder ($i=\mathrm{D}$). Assuming a large displaced state for both subsystems, we shift the bosonic modes as $\hat{b}_i \to \hat{b}_i + \sqrt{S}\beta_i$, where the $\beta_i$ are complex fields yet to be determined. As $S$ is a large displacement, we expand all operators in the small parameter $\epsilon=1/\sqrt{S}$
\begin{align}
    \hat{m}_{\alpha}^{i} = \frac{\hat{S}_{\alpha}^{i}}{S} = \sum_{l=0}^{\infty} \epsilon^l \hat{m}_{\alpha,l}^{i}.
    \label{eq:rescaled operators cascaded system}
\end{align}
Similar to the case of a single system, the first order expansion reads
\begin{align}
    \hat{m}_+^{i} \approx m_{+,0}^{i} + \underset{\hat{m}_{+,1}^{i}}{\underbrace{ \epsilon\frac{1}{2\sqrt{k_i}}[(4-3|\beta_i|^2) - |\beta_i|^2)\hat{b}_i^{\dagger} - \beta_i^{*2}\hat{b}_i]}},
\end{align}
where $k_i = 2-|\beta_i|^2$ and $m_{+,0}^{i} = \beta_i^{*} \sqrt{k_i}$. Inserting the rescaled operators and rescaled parameters \eqref{eq:rescaled parameters} in the Lindblad master equation for the cascaded system yields
\begin{align}
\begin{split}
    \partial_{\tau} \rho_\mathrm{ss} &= -iS\Tilde{\omega} [\hat{m}_\mathrm{x}^\mathrm{S} + \hat{m}_\mathrm{x}^\mathrm{D},\rho_\mathrm{ss}] \\
    &\quad\,- \frac{\kappa}{2}S [e^{-i\Delta \varphi}\hat{m}_+^\mathrm{D}\hat{m}_-^\mathrm{S} - e^{i\Delta\varphi}\hat{m}_+^\mathrm{S}\hat{m}_-^\mathrm{D}, \rho_\mathrm{ss}]\\
    &\quad\,+ \kappa S \mathcal{D}[e^{-i\varphi}\hat{m}_-^\mathrm{S} + e^{-i\varphi'}\hat{m}_-^\mathrm{D}]\rho_\mathrm{ss}\\
    &\overset{!}{=} 0,
    \label{cascaded-stationary-rescaled}
\end{split}
\end{align}
which we want to solve for the stationary state $\rho_\mathrm{ss}$. By expanding the density matrix $\rho = \sum_{l=0}^{\infty} \epsilon^l \rho_l$, we find up to order $\mathcal{O}(\epsilon)$ self-consistency relations
\begin{align}
    m_{\pm,0}^\mathrm{S} &= \pm \frac{i\Tilde{\omega}}{\kappa}, \hspace{0.2cm} m_{\pm,0}^\mathrm{D} = \pm \frac{i\Tilde{\omega}}{\kappa}(1-2e^{\pm i\Delta \varphi}).
    \label{eq:leading order ladder operators cascaded system}
\end{align}
With the solution for the zero order moments of the ladder operators, we determine the fields $\beta_i$ as
\begin{align}
\begin{split}
    \beta_\mathrm{S}&= -i \sqrt{1-\sqrt{1-(\Tilde{\omega}/\kappa)^2}}, \\
    \beta_\mathrm{D} &= -\frac{i\Tilde{\omega}}{\kappa}\frac{1-2e^{-i\Delta\varphi}}{\sqrt{1+\sqrt{1-(\Tilde{\omega}^2/\kappa^2 )(5-4\cos(\Delta \varphi))}}}.
\end{split}
\end{align}
This solution is only valid in the regime $\Tilde{\omega}^2 \leq \kappa^2/(5-4\cos(\Delta \varphi))$ or in other terms, we obtain an additional transition line for the cascaded system along $\omega_\mathrm{c,casc} = \omega_\mathrm{c}/\sqrt{5-4\cos(\Delta\varphi)}$, with the single system critical frequency $\omega_\mathrm{c}$. The next to leading order term yields
\begin{align}
\begin{split}
    \partial_{\tau} \rho_0 &= -\frac{\kappa}{2} [e^{-i\Delta \varphi}\hat{m}_{+,1}^\mathrm{D}\hat{m}_{-,1}^\mathrm{S} - e^{i\Delta \varphi}\hat{m}_{+,1}^\mathrm{S}\hat{m}_{-,1}^\mathrm{D}, \rho_0] \\
     &\;\;\;\;\;+ \kappa \mathcal{D}[e^{-i\varphi}\hat{m}_{-,1}^\mathrm{S} + e^{-i\varphi'}\hat{m}_{-,1}^\mathrm{D}]\rho_0 \overset{!}{=} 0.
\end{split}
\end{align}
We solve this by the ansatz
\begin{align}
    \rho_{0,\mathrm{ss}} = (\ket{E_0^\mathrm{S}}\otimes \ket{E_0^\mathrm{D}})(\bra{E_0^\mathrm{S}}\otimes\bra{E_0^\mathrm{D}}),
    \label{eq:leading order stationary state cascaded system}
\end{align}
with the vacuum of the fluctuations $\hat{m}_{-,1}^{i} \ket{E_0^{i}}=0$. By writing out the fluctuations as $\hat{m}_{-,1}^{i} = A_i \hat{b}_i + B_i \hat{b}_i^{\dagger}$ with
\begin{align}
    B_\mathrm{S} &= -\frac{1-\sqrt{1-\Tilde{\omega}^2/\kappa^2}}{2\sqrt{k_\mathrm{S}}}, \\
    A_\mathrm{S} &= \frac{1+3\sqrt{1-\Tilde{\omega}^2/\kappa^2}}{2\sqrt{k_\mathrm{S}}},\\
    B_\mathrm{D} &= \frac{(-i\Tilde{\omega}/\kappa)1-2e^{i\Delta\varphi}}{\sqrt{1+\sqrt{1-(\Tilde{\omega}^2/\kappa^2)(5-4\cos(\Delta \varphi))}}\sqrt{2k_\mathrm{D}}}, \\
    A_\mathrm{D} &= \frac{1+3\sqrt{1-(\Tilde{\omega}^2/\kappa^2)(5-4\cos(\Delta\varphi))}}{2\sqrt{k_\mathrm{D}}},
\end{align}
the explicit form of the stationary state is determined. Notice that $|B_i / A_i| \leq 1$ and thus we find the vacuum states of the respective fluctuations
\begin{align}
    \ket{E_0^{i}} = \frac{1}{\sqrt{\mathcal{N}}}\left( \ket{0}_i + \sum_{n=1}^{\infty} \left( \frac{-B_i}{A_i}\right)^n \sqrt{\frac{(2n-1)!!}{2n!!}} \ket{2n}_i \right). 
\end{align}

\begin{figure}[t]
    \centering
    \vspace{-0.3cm}
    \includegraphics{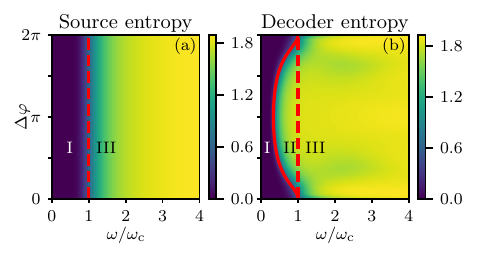}
    \vspace{-0.6cm}
    \mycaption{Stationary von-Neumann entropy of the cascaded system.}{
    $N=6$ for all panels. 
    Von-Neumann entropy of the stationary state reduced to (a) the source and (b) the decoder system varying Rabi frequency $\omega$ and phase difference $\Delta \varphi$.
    Red-solid lines: transition line $\omega_\mathrm{c}$  where both source and decoder display time crystal behavior (\MakeUppercase{\romannumeral 3}).
    Red-dashed line in (b): mean-field transition line $\omega_\mathrm{c, casc}$ from the stationary regime (\MakeUppercase{\romannumeral 1}) to the intermediate regime, where only the decoder is in a time crystal regime (\MakeUppercase{\romannumeral 2}).}
    \label{fig:S_VN stationary casc}
\end{figure}

\begin{figure}[t]
    \centering
    \vspace{-0.3cm}
    \includegraphics{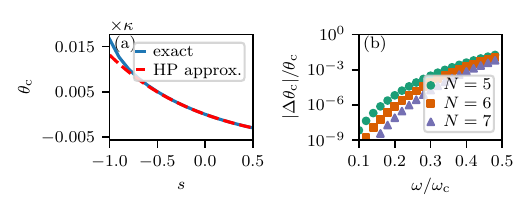}
    \vspace{-0.5cm}
    \mycaption{Benchmark of the HP approximation for the perfect absorber protocol.}{(a) Dominant eigenvalue of \eqref{eq:tilted me cascaded system} as a function of the bias parameter $s$, for $\Delta \varphi=0.1$, fixed system size $N=5$ and in the stationary regime $\omega/\omega_\mathrm{c}=0.35$. Comparison of results obtained with the HP approximation (red dashed line) and numerical diagonalization of \eqref{eq:tilted me cascaded system} (blue solid line). (b) Relative error of the dominant eigenvalue $\theta_\mathrm{c}$ as a function of the Rabi frequency $\omega$, for different system sizes, and fixed values $\Delta\varphi=0.1$ and $s=-0.05$. The error is defined as the absolute difference of the HP approximated value and the exact diagonalized value and both divided by the exact value. The error increases towards the critical Rabi frequency, but is small in a large portion of the stationary regime. The approximation improves with increasing system size.}
    \label{fig:hp approximation estimation error cascaded}
\end{figure}

\begin{figure*}[t]
    \centering
    \vspace{-0.3cm}
    \includegraphics{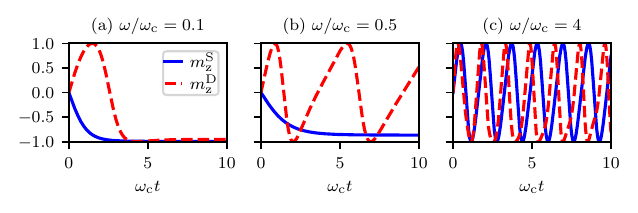}
    \vspace{-0.5cm}
    \mycaption{Cascaded system mean field quantities.}{Mean field expectation value of the $\mathrm{z}-$magnetization $\lim_{N\to\infty}(2/N) \langle\hat{S}_\mathrm{z}^i\rangle = m_\mathrm{z}^i $, with $i=\mathrm{S}$ for the source (blue solid line) and $i=\mathrm{D}$ the decoder (red dashed line), for different positions in the phase diagram. $\omega_\mathrm{c} = N\kappa/2$ is the critical frequency and both $\omega$ and $t$ are measured in units of $\omega_\mathrm{c}$ to obtain a well-defined limit. For all panels, $\Delta\varphi = \pi$. (a) $\omega/\omega_\mathrm{c}=0.1$, with source and decoder in the stationary regime. (b) $\omega/\omega_\mathrm{c}=0.5$, with the source in the stationary regime and the decoder displaying time crystal behavior. (c) $\omega/\omega_\mathrm{c}=4$, with both source and decoder in the time crystal regime.}
    \label{fig:cascaded meand field btc}
\end{figure*}

The additional transition line at $\omega = \omega_\mathrm{c,casc}$ is strongly visible in the von Neuman entropy \mbox{$S_\mathrm{VN} = -\Tr [\rho^i_\mathrm{ss} \ln \rho^i_\mathrm{ss}]$} of the stationary state in the perfect absorber protocol reduced to the source (\mbox{$i=\mathrm{S}$}) and the decoder (\mbox{$i=\mathrm{D}$}). For the source, a transition from a pure to a mixed stationary state occurs at \mbox{$\omega=\omega_\mathrm{c}$}, whereas for the decoder this transition is at \mbox{$\omega=\omega_\mathrm{c,casc}$} [see \figref{fig:S_VN stationary casc}]. This shows that the transition at a driving strength \mbox{$\omega = \omega_\mathrm{c,casc}$} observed in the stationary purity and intensity with respect to the global state of the cascaded system in Fig. 3 of the main text are due to a transition of the decoder system. For driving strengths below $\omega_\mathrm{c}$, the source system remains in a pure stationary state, which underlines the causal structure of the cascaded system: photons, once emitted by the source are not fed back to the source and therefore only affect the decoder and its output. Therefore, the stationary state of the source is oblivious of any changes in the dynamics of the decoder. The decoder, however, receives the source output as an additional input, leading to a change in its stationary state and transitioning to a time crystal regime for lower values of $\omega$. In the strong-driving limit $\omega \to \infty$, we observe that the entropy of both subsystems approaches its maximum value, indicating that source and decoder are in a highly entangled state.

\subsection{Estimation error for the perfect absorber protocol in the stationary regime}

Inserting the rescaled parameters \eqref{eq:rescaled parameters} and operators \eqref{eq:rescaled operators cascaded system} in the tilted master equation \eqref{eq:tilted me cascaded system} yields
\begin{widetext}
\begin{align}
    \partial_\tau \rho = \mathcal{L}[\rho] + \kappa S (e^{-s} -1)(e^{-i\varphi} \hat{m}_-^\mathrm{S} + e^{-i\varphi'}\hat{m}_-^\mathrm{D})\rho(e^{i\varphi} \hat{m}_+^\mathrm{S} + e^{i\varphi'}\hat{m}_+^\mathrm{D}).
\end{align}
\end{widetext}
For the dominant eigenstate we choose the ansatz as in \eqref{eq:leading order stationary state cascaded system}, $\rho_{0,\mathrm{ss}} = (\ket{E_0^\mathrm{S}}\otimes \ket{E_0^\mathrm{D}})(\bra{E_0^\mathrm{S}}\otimes\bra{E_0^\mathrm{D}})$. Inserting this ansatz on the right hand side and only considering the leading terms in the expansion of the operators yields the dominant eigenvalue
\begin{widetext}
\begin{align}
\begin{split}
    \theta_\mathrm{c} (s,\Delta\varphi) &\approx \kappa S^2 (e^{-s}-1)(m_{-,0}^\mathrm{S}m_{+,0}^\mathrm{S} + m_{-,0}^{\mathrm{D}}m_{+,0}^{\mathrm{D}}+e^{-i\Delta \varphi}m_{-,0}^{\mathrm{S}}m_{+,0}^{\mathrm{D}} + e^{i\Delta \varphi}m_{-,0}^{\mathrm{D}}m_{+,0}^{\mathrm{S}} \\
    &=S^2(e^{-s} -1)\frac{\Tilde{\omega}^2}{\kappa}2(1-\cos(\Delta \varphi)) =2(e^{-s}-1)\frac{\omega^2}{\kappa}(1-\cos(\Delta \varphi)),
\end{split}
\end{align}
\end{widetext}
which is benchmarked with data obtained by diagonalization of \eqref{eq:tilted me cascaded system} in \figref{fig:hp approximation estimation error cascaded}. From this, the estimation error is computed as
\begin{align}
    \overline{\delta \varphi} = \frac{\sqrt{\partial_s^2 \theta_\mathrm{c} (s,\Delta\varphi)}}{|\partial_{\varphi}\partial_s\theta_\mathrm{c} (s,\Delta\varphi)|}\big |_{s=0} = \frac{\sqrt{\kappa(1-\cos(\Delta \varphi))}}{\sqrt{2}\omega|\sin(\Delta \varphi)|},
\end{align}
which is minimal and saturates the QCRB for $\Delta\varphi = 0$. Note that this approximation only holds in the cascaded stationary regime for $\omega < \omega_\mathrm{c,casc}$.

\subsection{Mean-field operators for the source-decoder system}

We define the rescaled operators as $\hat{m}_{\alpha}^{i,N} = 2\hat{S}^i_{\alpha}/N$, with $\alpha \in \{\mathrm{x,y,z}\}$ and ($i=\mathrm{S}$) for the source and ($i=\mathrm{D}$) for the decoder. In the limit $N\to\infty$, these operators converge to multiples of the identity, i.e. $\lim_{N\to\infty}\hat{m}_{\alpha}^{i,N} =\lim_{N\to\infty}\langle\hat{m}^{i,N}_{\alpha}\rangle =:m^i_{\alpha}$ \cite{Carollo2022, Benatti2018, Lanford1969}. As in the main text we consider the maximally polarized sector with total angular momentum of the subsystems $S=N/2$. While the time evolution of the density matrix is given by a Lindblad master equation, the evolution of expectation values is given by an adjoint Lindblad master equation
\begin{widetext}
\begin{align}
\begin{split}
    \partial_t \langle\hat{m}_{\alpha}^N\rangle= &-i\omega \biggr\langle \left[\hat{m}_{\alpha}^N, \hat{S}_\mathrm{x}^{\mathrm{S}}+\hat{S}_\mathrm{x}^{\mathrm{D}}\right]\biggr\rangle - \frac{\kappa}{2} \biggr\langle\left[\hat{m}_{\alpha}^N,e^{-i\Delta \varphi}\hat{S}_+^{\mathrm{D}}\hat{S}_-^{\mathrm{S}} - e^{i\Delta\varphi}\hat{S}_+^{\mathrm{S}}\hat{S}_-^{\mathrm{D}}\right]\biggr\rangle\\
    &+\kappa\biggr\langle\left(e^{i\varphi}\hat{S}_+^{\mathrm{S}}+e^{i\varphi'}\hat{S}_+^{\mathrm{D}}\right)\hat{m}_{\alpha}^N \left(e^{-i\varphi}\hat{S}_-^{\mathrm{S}} + e^{-i\varphi'}\hat{S}_-^{\mathrm{D}}\right)\biggr\rangle \\
    &- \frac{\kappa}{2}\biggr\langle \left\{\left(e^{i\varphi}\hat{S}_+^{\mathrm{S}}+e^{i\varphi'}\hat{S}_+^{\mathrm{D}}\right)\left(e^{-i\varphi}\hat{S}_-^{\mathrm{S}} + e^{-i\varphi'}\hat{S}_-^{\mathrm{D}}\right),\hat{m}_{\alpha}^N\right\}\biggr\rangle.
\end{split}
\end{align}
\end{widetext}
In a first step, this yields the equations of motion for the expectation values $\langle \hat{m}_\alpha^{i,N}\rangle$. We then continue by defining $\omega_\mathrm{c} = N\kappa/2$, and rescaling time and the Rabi frequency as $\tau = St, \omega = S\Tilde{\omega}$. With this rescaling, we then take the thermodynamic limit, assuming that expectation values factor as $\langle \hat{m}_\alpha^{i,N} \hat{m}_\beta^{j,N} \rangle \overset{N\to\infty}{\longrightarrow} \langle \hat{m}_\alpha^{i,N}\rangle \langle \hat{m}_\beta^{j,N} \rangle$. With this we find the equations of motion 
\begin{widetext}
\begin{align}
    \partial_\tau m_\mathrm{x}^{\mathrm{S}} &= \kappa m_\mathrm{x}^{\mathrm{S}} m_\mathrm{z}^{\mathrm{S}}     \label{eq:mean field equations of motion cascaded system 1}\\
    \partial_\tau m_\mathrm{y}^{\mathrm{S}} &= -\Tilde{\omega} m_\mathrm{z}^{\mathrm{S}} + \kappa m_\mathrm{y}^{\mathrm{S}} m_\mathrm{z}^{\mathrm{S}} \\
    \partial_\tau m_\mathrm{z}^{\mathrm{S}} &= \Tilde{\omega} m_\mathrm{y}^{\mathrm{S}} - \kappa   \left[(m_\mathrm{x}^{\mathrm{S}})^2 + (m_\mathrm{y}^{\mathrm{S}})^2\right]\\
    \partial_\tau m_\mathrm{x}^{\mathrm{D}} &= \kappa \left[m_\mathrm{x}^{\mathrm{D}}m_\mathrm{z}^{\mathrm{D}} + 2m_\mathrm{x}^{\mathrm{S}}m_\mathrm{z}^{\mathrm{D}}\cos (\Delta \varphi)-2m_\mathrm{y}^{\mathrm{S}}m_\mathrm{z}^{\mathrm{D}}\sin (\Delta \varphi)\right]\\
    \partial_\tau m_\mathrm{y}^{\mathrm{D}} &= -\Tilde{\omega} m_\mathrm{z}^{\mathrm{D}} + \kappa \left[m_\mathrm{z}^{\mathrm{D}}m_\mathrm{y}^{\mathrm{D}} + 2m_\mathrm{x}^{\mathrm{S}}m_\mathrm{z}^{\mathrm{D}}\sin (\Delta\varphi)+2m_\mathrm{y}^{\mathrm{S}}m_\mathrm{z}^{\mathrm{D}}\cos(\Delta\varphi)\right]\\
    \begin{split}
    \partial_\tau m_\mathrm{z}^{\mathrm{D}} &= \Tilde{\omega} m_\mathrm{y}^{\mathrm{D}}-2\kappa \left[\left(m_\mathrm{x}^{\mathrm{S}}m_\mathrm{x}^{\mathrm{D}} + m_\mathrm{y}^{\mathrm{S}}m_\mathrm{y}^{\mathrm{D}}\right)\cos (\Delta\varphi)  + \left(m_\mathrm{x}^{\mathrm{S}}m_\mathrm{y}^{\mathrm{D}} - m_\mathrm{y}^{\mathrm{S}}m_\mathrm{x}^{\mathrm{D}}\right)\sin (\Delta\varphi))\right]- \kappa \left[(m_\mathrm{x}^{\mathrm{D}})^2 + (m_\mathrm{y}^{\mathrm{D}})^2\right].
    \end{split}
    \label{eq:mean field equations of motion cascaded system 2}
\end{align}
\end{widetext}
The operators $(\hat{m}^i)^2$ commute with all components of the angular momentum, meaning that the expectation values $\langle (\hat{m}^i)^2  \rangle$ are conserved, and due to the maximum polarization condition $(\hat{S}^i)^2 = N/2$, we find $(m^i)^2 = 1$. Using this boundary condition, we numerically compute the dynamics generated by \multieqref{eq:mean field equations of motion cascaded system 1}{eq:mean field equations of motion cascaded system 2} for three different ratios $\Tilde{\omega}/\kappa$ and display the results for the $m_\mathrm{z}^i$ components in \figref{fig:cascaded meand field btc}. We further determine the values of $m_{\pm}^{i}$ in the thermodynamic limit and in the stationary state. By setting \multieqref{eq:mean field equations of motion cascaded system 1}{eq:mean field equations of motion cascaded system 2} to zero, we determine the stationary solutions for the mean field observables as 
\begin{align}
    m_{\pm}^{\mathrm{S}} &= \pm i\frac{\Tilde{\omega}}{\kappa}, \\
    m_{\pm}^{\mathrm{D}} &= \pm i\frac{\Tilde{\omega}}{\kappa} - 2m_\pm^\mathrm{S}e^{\pm i\Delta\varphi} = \pm i\frac{\Tilde{\omega}}{\kappa}(1-2e^{\pm i \Delta \varphi}),
\end{align}     
which agrees with the results obtained with the HP approach for $m_{\pm,0}^i$ in \eqref{eq:leading order ladder operators cascaded system}.



\begin{figure*}[]
    \centering
    \vspace{-0.3cm}
    \includegraphics[width=0.95\textwidth]{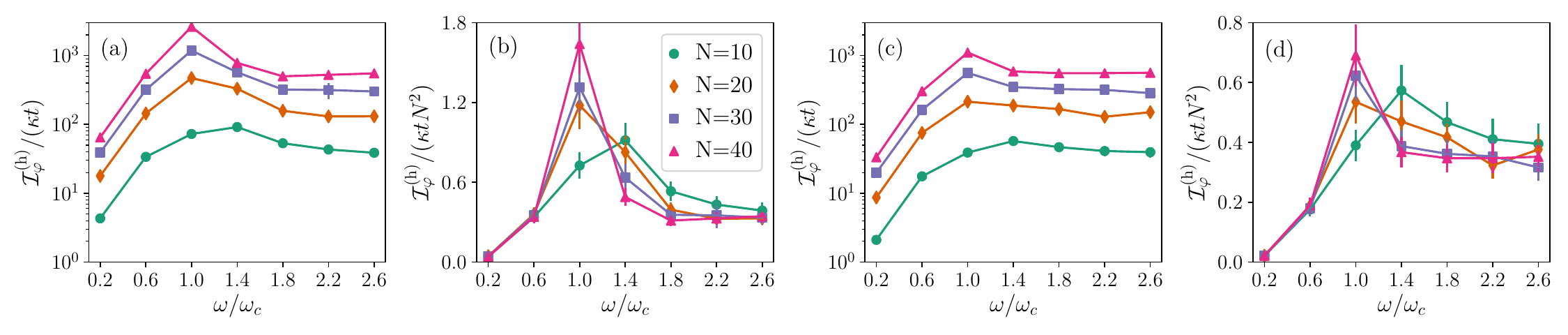}
    \vspace{-0.5cm}
    \mycaption{Long-time FI rate of the homodyne current.}{Parameter values: (a), (b) $\varphi-\beta =0$; (c), (d) $\varphi-\beta =\pi/4$. The FI rates have been obtained by sampling $1000$ trajectories. Error bars represent the Monte‑Carlo errors multiplied by a factor of 3. The results have been also averaged over a time window in which the FI rate had already reached its stationary value, specifically from from $\kappa t=15$ to $\kappa t=40$.}
    \label{fig:fi_homodyne}
\end{figure*}

\section{Fisher Information of the homodyne monitoring record and the perfect absorber protocol}\label{SecSM:classical_FI}

In this work, we also analyze the Fisher information associated with the homodyne measurement record and with the photocounting record \cite{gammelmark_bayesian_2013,Kiilerich2016,Albarelli2018,Mattes2025}. These quantities are introduced in detail in Refs. \cite{gammelmark_bayesian_2013,Albarelli2018}, where various algorithms for their computation are presented. Here, we adopt the method of Ref. \cite{Albarelli2018}, which provides an efficient algorithm based on the Kraus-operator description of ideal homodyne detection and ideal photocounting. Below we present a brief introduction to the main ingredients required to compute and interpret this quantity. For a more detailed treatment, we refer the reader to the aforementioned references.

\subsection{Method of computation for ideal homodyne monitoring}

For our system, assuming unit detection efficiency and discretizing time into small steps $\delta t$, homodyne trajectories can be generated using the following Kraus operators. At time $t$, the Kraus operator conditioned on the homodyne current $J_t$ is
\begin{equation}
K_{J_t} = \mathds{1}-i\delta t \Omega \hat{S}_\mathrm{x} - \frac{\kappa \delta t}{2} \hat{S}_+ \hat{S}_- + \sqrt{\kappa} e^{i(\beta-\varphi)} \hat{S}_- {J_t}.
\end{equation}
The homodyne current is sampled according to:
\begin{equation}
J_t=\langle \psi_{t-\delta t}| (S_- e^{i(\beta-\varphi)}+S_+ e^{-i(\beta-\varphi)})|  \psi_{t-\delta t} \rangle \delta t  +\delta W_t  
\end{equation}
where $\delta W_t$ is a Wiener increment at time $t$ and   $|  \psi_{t-\delta t} \rangle$ is the conditional state for the given homodyne trajectory at the previous time step \cite{WisemanMilburn2009,Albarelli2018,Mattes2025}. The angle $\beta$ denotes the chosen homodyne phase.

Assuming a pure initial state, the unnormalized conditional state for a given homodyne trajectory at time $t$ is given by:
\begin{equation}
|\tilde{\psi}_{t}\rangle=K_{J_t} K_{J_{t-\delta t}} \dots K_{J_{\delta t}}|\psi_{0}\rangle.
\end{equation}
The full homodyne measurement record is the set:
\begin{equation}
\mathbf{J_t}=\{J_{\delta t},\dots, J_{t-\delta t},J_{t}\} ,  
\end{equation}
which uniquely specifies the quantum trajectory. Each record occurs with a probability  $P(\mathbf{J_t}|\varphi)$, that depends on the parameter of interest $\varphi$. This probability can be computed from the unnormalized conditional state  \cite{WisemanMilburn2009,Albarelli2018,Mattes2025}.

The Fisher information associated with the homodyne monitoring record up to time $t$ for estimating $\varphi$ around $\varphi_0$ is defined as $\mathcal{I}_\varphi^\mathrm{(h)}(\varphi_0,t)=\mathbb{E}\big[\big(\partial_\varphi \log P(\mathbf{J_t}|\varphi_0 )\big)^2\big]$. The ensemble average $\mathbb{E}$ can be evaluated using the Monte‑Carlo method of Ref. \cite{Albarelli2018}: one generates many homodyne trajectories, computes the Fisher information contribution for each trajectory, and finally averages over the ensemble. Following the method of Ref. \cite{Albarelli2018}, the Fisher information can be expressed as:
\begin{equation}
\mathcal{I}_\varphi^\mathrm{(h)}(\varphi_0,t)=\mathbb{E}\big[\big( \langle \phi_t|\psi_t\rangle +\langle \psi_t|\phi_t \rangle\big)^2\big],
\end{equation}
where the auxiliary vector $|\phi_t\rangle$ is propagated alongside the conditional state $|\psi_t\rangle$ for each homodyne realization. Its evolution is given by
\begin{equation}
|\phi_t\rangle=\frac{(\partial_\varphi K_{J_t})|\psi_{t-\delta t}\rangle+K_{J_t}|\phi_{t-\delta t}\rangle}{\sqrt{\langle \psi_{t-\delta t}| K_{J_t}^\dagger K_{J_t}|\psi_{t-\delta t}\rangle}}.    
\end{equation}

\subsection{Method of computation for ideal photocounting}

Similarly, assuming unit detection efficiency and discretizing time as before, photocounting trajectories for the absorber protocol are generated using the Kraus operators
\begin{align}
    K_0 &= \mathds{1} - i\delta t \left[ \omega \left( \hat{S}_\mathrm{x}^\mathrm{S} + \hat{S}_\mathrm{x}^\mathrm{D} \right) + \hat{H}_\mathrm{casc}(\Delta \varphi) \right]\\
    &\;\;\;\;\;- \frac{\kappa \delta t}{2} \hat{L}_\mathrm{casc}^\dagger(\Delta \varphi) \hat{L}_\mathrm{casc}(\Delta \varphi), \\
    K_1 &= \sqrt{\kappa \delta t} \hat{L}_\mathrm{casc}(\Delta \varphi),
\end{align}
with the cascaded Hamiltonian $\hat{H}_\mathrm{casc}(\Delta \varphi) = - (i\kappa /2) (e^{-i\Delta\varphi} \hat{S}_+^\mathrm{D}\hat{S}_-^\mathrm{S} - e^{i\Delta\varphi} \hat{S}_+^\mathrm{S}\hat{S}_-^\mathrm{D})$, and cascaded jump operator $\hat{L}_\mathrm{casc}(\Delta \varphi) = e^{-i\Delta\varphi} \hat{S}_-^\mathrm{S} + \hat{S}_-^\mathrm{D}$ depending only on the phase difference $\Delta \varphi = \varphi - \varphi'$. Assuming pure initial states, the state at time $t$ reads
\begin{align}
    |\Tilde{\psi}_t\rangle = K_{i_t} K_{i_{t-\delta t}} \cdots K_{i_{\delta t}} \ket{\psi_0},
\end{align}
where the index $i_t = 0,1$ indicates whether a photon is detected at time $t$. The full record of emissions is the set of indices
\begin{align}
    \mathbf{I_t} = \{ i_{\delta t} , \dots , i_{t - \delta t} ,i_t\},
\end{align}
which occurs with probability $P(\mathbf I_t |\varphi)$, that is conditional on the parameter of interest $\varphi$. This probability is computed from the unnormalized conditional state.
As in the case of homodyne detection, the Fisher information associated with the record of emissions up to time $t$ for estimating $\varphi$ around $\varphi_0$ is defined as $\mathcal{I}^\mathrm{(c)}_\varphi (\varphi_0 , t) = \mathbb{E}[\left(\partial_\varphi \log P(\mathbf I_t | \varphi_0) \right)^2]$, where $\mathbb{E}$ is the ensemble average over trajectories, which we evaluate using the method of Ref. \cite{Albarelli2018}. Following the derivation of Ref. \cite{Albarelli2018}, the Fisher information simplifies to
\begin{align}
    \mathcal{I}^\mathrm{(c)}_\varphi (\varphi_0 , t) = \mathbb{E}[\left( \braket{\phi_t}{\psi_t} + \braket{\psi_t}{\phi_t}\right)^2],
\end{align}
where the auxiliary vector $\ket{\phi_t}$ is defined as before but replacing the Kraus operators $K_{J_t}$ by the photocounting Kraus operators.

\begin{figure*}[b]
    \centering
    \includegraphics[]{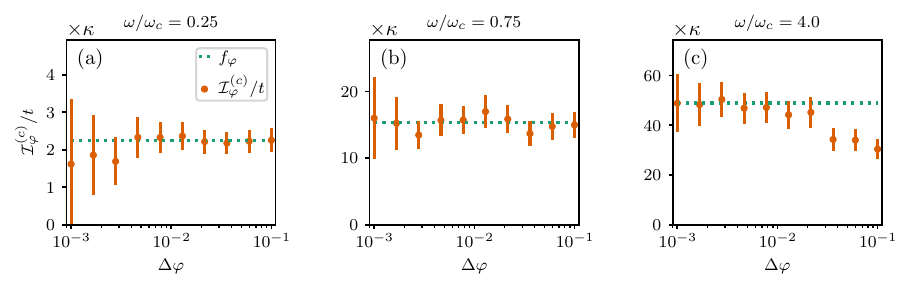}
    \vspace{-0.5cm}
    \mycaption{FI rate of the photocounting signal in the perfect absorber protocol.}{Parameter values: All panels $N=6$, (a) $\omega / \omega_\mathrm{c} = 0.25$, (b) $\omega / \omega_\mathrm{c} = 0.75$, and (c) $\omega / \omega_\mathrm{c} = 4$. The FI rates are obtained by sampling $10^3$ trajectories up to a time of $\kappa t = 3 \times 10^4$ and averaging the FI rate over the time window $\kappa t \in [0.5\times 10^4 , 3\times 10^4]$. The error bars represent three times the standard deviation of the Monte-Carlo sampling. The green dotted line indicates the long-time QFI rate computed using exact diagonalization.}
    \label{fig:fi_perfect_absorber}
\end{figure*}

\begin{figure*}[t]
    \centering
    \vspace{-0.5cm}
    \includegraphics[]{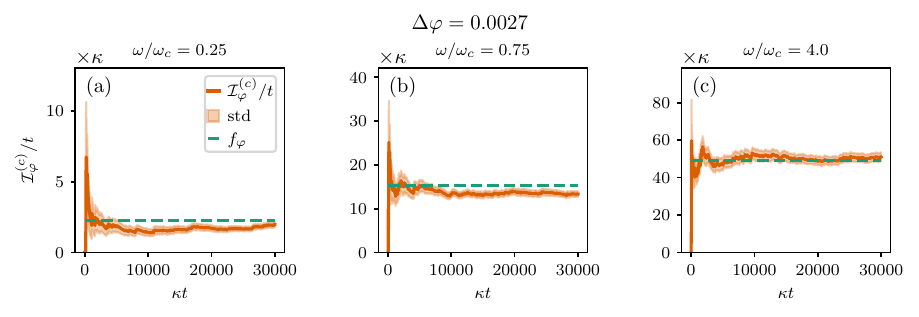}
    \vspace{-0.5cm}
    \mycaption{Time resolved FI rate of the photocounting signal in the perfect absorber protocol at $\Delta \varphi=0.0027$.}{Parameter values: All panels $N=6$ and $\Delta \varphi = 0.0027$. (a) $\omega / \omega_\mathrm{c} = 0.25$, (b) $\omega / \omega_\mathrm{c} = 0.25$, and (c) $\omega / \omega_\mathrm{c} = 4$. The FI rates (orange solid line) are obtained by sampling $10^3$ trajectories up to a time of $\kappa t = 3\times 10^4$. The orange shaded area represents the standard deviation of the Monte-Carlo sampling over trajectories. The green dashed line indicates the long-time QFI rate obtained using exact diagonalization.}
    \label{fig:FI_trajs_vs_qfi_dphi0.0027}
\end{figure*}

\begin{figure*}[t]
    \centering
    \vspace{-0.5cm}
    \includegraphics[]{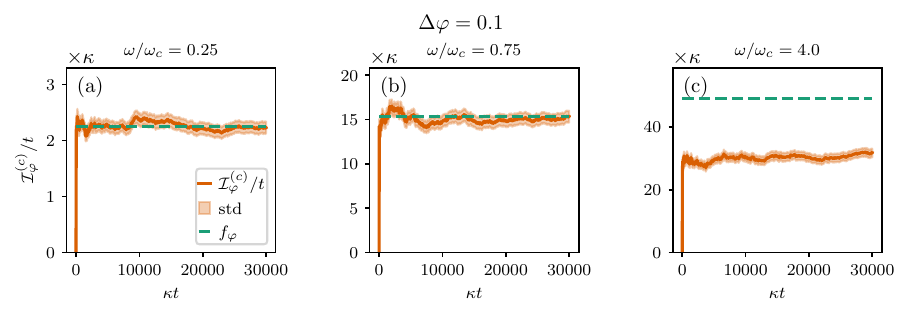}
    \vspace{-0.5cm}
    \mycaption{Time resolved FI rate of the photocounting signal in the perfect absorber protocol at $\Delta \varphi=0.1$.}{Parameter values: All panels $N=6$ and $\Delta \varphi = 0.1$. (a) $\omega / \omega_\mathrm{c} = 0.25$, (b) $\omega / \omega_\mathrm{c} = 0.25$, and (c) $\omega / \omega_\mathrm{c} = 4$. The FI rates (orange solid line) are obtained by sampling $10^3$ trajectories up to a time of $\kappa t = 3\times 10^4$. The orange shaded area represents the standard deviation of the Monte-Carlo sampling over trajectories. The green dashed line indicates the long-time QFI rate obtained using exact diagonalization.}
    \label{fig:FI_trajs_vs_qfi_dphi0.1}
\end{figure*}

\subsection{Results for homodyne detection}

As explained in the main text, the FI eventually reaches a long‑time regime in which it becomes proportional to $t$, i.e. $\mathcal{I}_\varphi^{(\mathrm{h})}(\varphi_0,t)\propto t$ \cite{gammelmark_bayesian_2013,Kiilerich2016,Albarelli2018,Mattes2025}. The time required to enter this regime shows some dependence on $\omega/\omega_\mathrm{c}$, but only a weak dependence on $N$. In practice, for times of order $\kappa T\sim 10$, the FI has already converged to this stationary behavior. We therefore focus on this long‑time regime and examine whether the rate $\mathcal{I}_\varphi^{(\mathrm{h})}$ displays the same scaling as the QFI.

In Figure \ref{fig:fi_homodyne} we show the results for $\beta-\varphi=0$ [panels (a),(b)] and $\beta-\varphi=\pi/4$ [panels (c),(d)]. We also checked $\beta-\varphi=\pi/3$, obtaining qualitatively similar behavior. We consider system sizes from $N=10$ to $N=40$. In general, larger $N$ and larger $\omega/\omega_\mathrm{c}$ require smaller $\delta t$ to ensure convergence. Each data point corresponds to an average over 1000 trajectories of total duration $\kappa t=40$. For each trajectory, the FI rate is obtained by time‑averaging between $\kappa t=15$ and $\kappa t=40$. Error bars represent three times the Monte‑Carlo error.

Panels (a) and (c) display the bare FI rate, while panels (b) and (d) show the same data rescaled by $N^2$. In the stationary regime, the homodyne FI rate saturates the QFI rate for the corresponding parameter values. This is more clearly visible in Fig. 2 of the main text (for $\beta-\varphi=0$), where we plot the inverse square root of the FI rate together with the estimation error of the average‑current protocol, which also saturates the QFI in this regime. The FI reaches its maximum at the phase transition and then decreases to values comparable to those in the stationary phase. As with the QFI rate, for sufficiently large $\omega/\omega_\mathrm{c}$, the FI rate shows little dependence on this parameter.

The most important observation appears in panels (b) and (d): in the time crystal regime, the FI rates for different system sizes collapse when rescaled by $N^2$. This indicates that the FI grows as $N^2$ in this regime, in contrast with the $N^4$ scaling of the QFI. Although the accessible system sizes are modest and we have not explored the full range of $\beta-\varphi$, the result is significant: for the QFI, the $N^4$ scaling is already clearly visible within this same size range.

\subsection{Results for perfect absorber protocol}

As for the homodyne FI, the FI for photocounting in the perfect absorber protocol eventually reaches a long-time regime in which it becomes proportional to $t$, i.e. $\mathcal{I}^\mathrm{(c)}_\varphi \propto t$ \cite{gammelmark_bayesian_2013,Kiilerich2016,Albarelli2018,Mattes2025}. In other words, the long-time FI rate becomes constant, which we show in Figures \ref{fig:FI_trajs_vs_qfi_dphi0.0027} and \ref{fig:FI_trajs_vs_qfi_dphi0.1} both in the stationary regime for $\omega/\omega_\mathrm{c}=0.25$ in panel (a)  and $\omega / \omega_\mathrm{c} = 0.75$ in panel (b), and in the time crystal regime for $\omega / \omega_\mathrm{c}=4$ in panel (c) for a system size of $N=6$.

In Figure \ref{fig:fi_perfect_absorber}, we show the results for the FI rate for a range of phase differences $\Delta\varphi = 0.001,\dots,0.1$  both in the stationary regime [panels (a) and (b)] and in the time crystal regime [panel (c)]. Note, that these results are averaged over a time window of $\kappa t \in [0.5\times 10^4, 3\times 10^4]$, in which the cascaded system is fully relaxed to its stationary state after an initial relaxation phase. The error bars in this plot represent three times the Monte-Carlo error with respect to the sampled trajectories.

In the time crystal regime [panel (c) of Figure \ref{fig:fi_perfect_absorber}], the study of the long-time FI rate reveals that photocounting in the perfect absorber protocol allows for an optimal measurement of the phase shift $\varphi$, as at its maximum with respect to the phase difference $\Delta \varphi$, the FI rate saturates the QFI. However, this is only the case at an optimal value of $\Delta \varphi$, and for values of the phase difference away from this optimal point of operation the FI rate rapidly decreases. Note in particular that this optimal point is close to, but not precisely at the dark state condition. In addition, the FI rate decreases for larger values of $\Delta \varphi$ where the emitted intensity  increases rapidly. In Figures \ref{fig:FI_trajs_vs_qfi_dphi0.0027} and \ref{fig:FI_trajs_vs_qfi_dphi0.1}, we show the time resolved FI rate for two points in the phase diagram with $\Delta \varphi = 0.0027$ in Fig. \ref{fig:FI_trajs_vs_qfi_dphi0.0027} and $\Delta \varphi = 0.1$ in Fig. \ref{fig:FI_trajs_vs_qfi_dphi0.1}. These plots highlight, that in the stationary regime the perfect absorber protocol efficiently exploits the QFI for a broad range of values $\Delta\varphi$. In the time crystal regime, the FI for photocounting in the perfect absorber protocol also approaches the long-time QFI rate for long measurement times, when operated close to the optimal point and drops drastically when operated away from the optimal point.

Importantly, this shows that photocounting within the perfect absorber protocol enables optimal estimation, provided that one also chooses an appropriate estimator capable of extracting the full information contained in the photocounting record. In the main text we study the example of estimators based on the time averaged intensity. We find there that these are not optimal for the time crystal phase, in the sense that they do not saturate the QFI rate. However, our results demonstrate that, with suitable estimators (e.g., the pattern‑count estimator or maximum‑likelihood estimation \cite{Yang, Godley2023, Girotti2024}), the perfect absorber protocol achieves optimality, as the FI rate associated with photocounting saturates the QFI rate. Exploring these more complex estimation strategies constitutes an interesting direction for future work.


\bibliography{01_TeX/2nd_round/biblio}

\end{document}